\documentclass[%
 reprint,
 amsmath,amssymb,
 aps,
]{revtex4-2}

\usepackage{graphicx}
\usepackage{xcolor}
\usepackage{dcolumn}
\usepackage{bm}

\begin{document}

\preprint{APS/123-QED}

\title{Topologically-Protected Remanent Vortices in Confined Superfluid $^3$He}

\author{Alexander J. Shook}
\email{ashook@ualberta.ca}
\author{Daksh Malhotra}%
\author{Aymar Muhikira}%
\author{Vaisakh Vadakkumbatt}%
\author{John P. Davis}%
\email{jdavis@ualberta.ca}
\affiliation{%
 Department of Physics, University of Alberta, Edmonton, Alberta T6G 2E1, Canada}%

\date{\today}

\begin{abstract}
Thermodynamic phase transitions typically involve a transition in the microscopic ordering between a less and a more ordered state. After transitioning into an ordered state, localized structures known as defects may remain, which disrupt the overall macroscopic order. Kibble-Zurek theory predicts that for any second-order phase transition the density of defects that form should be determined by the scaling law for the system coherence time and the phase transition quench time. We have performed measurements of  sound dissipation due to vortex mutual friction in thin channels of superfluid $^3$He where one spatial dimension is smaller than a characteristic length scale predicted by the Kibble-Zurek theory. In this regime, we find that the density of defects is not determined by quench time, but can be estimated by replacing the Kibble-Zurek length scale with a truncated length set by the channel size. This results in significantly higher defect densities than in a bulk system.

\end{abstract}

\maketitle

\section{Introduction}

Phase transitions, the processes by which the microscopic order of a system spontaneously changes, are of fundamental importance to many scientific fields, not only in chemistry and condensed matter physics \cite{Papon2002} where the concept finds its origins, but also in cosmology \cite{Linde1979}, cell biology \cite{Pollack2008}, and economics \cite{Stanley2002}. 
One may understand phase transitions in terms of an order parameter, which quantifies the degree to which the system exhibits long-range correlations in a physical quantity specific to the system (e.g.,~density, magnetic moment, etc.) \cite{landau2013statistical,burmistrov2025statistical}. Such classification naturally gives rise to a length scale known as the correlation length, over which the system is ordered. A general feature of second-order phase transitions is the divergence of the system's correlation length at a critical temperature, representing the change from a disordered system (correlated only at very small distances) to one with long-range order. 


Correlation cannot spread throughout a system instantaneously. Shortly after the system crosses the critical temperature, the equilibrium correlation length is limited by the speed at which information propagates through the system. In a discussion of phase transitions in the early universe, Kibble \cite{Kibble1976} noted that correlation in a field, such as the scalar Higgs field, must be limited by causality since information cannot propagate faster than the speed of light. This implied the existence of cosmic domains where field correlations may only exist within the domain. Because each domain develops independently from every other, the expectation value of the field in each domain will be independent. Correlation in the Higgs field spreads as these causally disconnected domains propagate outward at the speed of light. As the field evolves in time, the topology of the initial domain structure may be such that topological defects, such as domain walls, cosmic strings, or monopoles persist.

\begin{figure}[t]
    \centering
    \includegraphics[width=\linewidth]{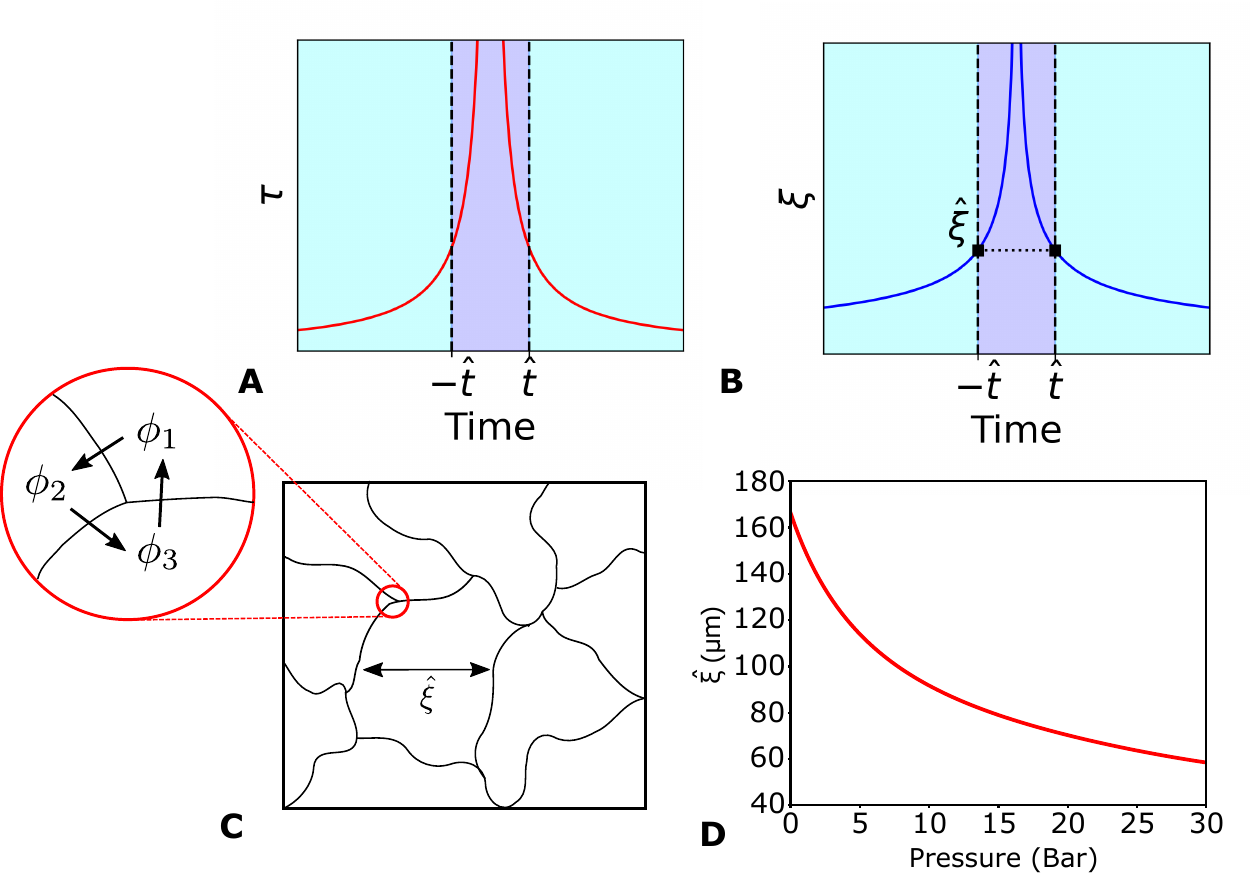}
    \caption{\textbf{Domain Illustration}. \textbf{\textbf{(a)}, \textbf{(b)} Scaling of coherence time, $\tau$, and length, $\xi$, respectively, as a function of time. Here, it is assumed that the temperature is changed linearly as a function of time. Both quantities diverge when the critical temperature is reached. The freeze-out time $\hat{t}$ is indicated by the dashed line. This time is defined as the moment when $\tau(\hat{t}) = \tau_Q = T_c(dT/dt)^{-1}$. At this moment, there is insufficient time for the domains to grow significantly, so the coherence length is assumed to be ``frozen" to first order at the value $\hat{\xi}$. (c)} During a rapid quench, domains with average size $\hat{\xi}$ form, giving rise to defects. Each domain will independently select a random phase $\phi$. If the phases $\phi_{1,2,3}$ increase sequentially, with a total change in phase close to $2\pi$ (clockwise or counter-clockwise), the resulting phase winding at the vertex will form a vortex. The density of vortices predicted to form is of order $\hat{\xi}^{-2}$. \textbf{(d)} The average domain size in $^3$He for a temperature ramp rate of 0.12 mK/hour is predicted by KZ theory to vary between 60 and 160 $\mu$m as a function of pressure \cite{SI}.}
    \label{fig:Domain_Cartoon}
\end{figure}

This argument was extended by Zurek, who discussed a similar process in the context of the phase transition from from normal to superfluid in liquid $^4$He. In this transition, the condensate wave function (i.e., the wavefunction of $^4$He atoms occupying the quantum ground state) becomes correlated at long distances \cite{Zurek1985}. Zurek took a new approach to estimating the density of defects by comparing the power-law divergence of the coherence time to the rate at which the system temperature is swept through the critical temperature (see Figure~\ref{fig:Domain_Cartoon}). 
If the system temperature or pressure is swept through a critical value quickly, then these domains will have insufficient time to propagate, leading to the creation of string-like defects known as quantized vortices.

This Kibble-Zurek mechanism (KZM) was generalized to describe a wide variety of phase transitions. Kibble-Zurek (KZ) theory describes this process in terms of an equilibrium correlation length $\xi$ and an equilibrium relaxation time $\tau$, both of which diverge at a second-order phase transition (see Figure 1 a and b). In the disordered phase the order parameter has a mean value of zero, but fluctuations exist that represent correlated regions of size $\xi$ that exist on a timescale of $\tau$. If a thermodynamic control parameter (such as temperature) is swept through the critical value where $\tau$ diverges, then there is a time after the phase transition is reached, known as the ``freeze-out time". This is the time at which correlated regions become long-lived relative to the amount of time required to reach equilibrium. This freeze-out time is therefore a function of the rate at which the control parameter is varied, as well as the temperature scaling of $\tau$. KZ theory predicts that the value of $\xi$ at freeze-out time determines the size of the independent order parameter domains in the ordered phase. Afterwards, there is a secondary process known as coarsening, where domains continue to grow in the ordered phase.




Each domain has a different microscopic structure, described by its order parameter, which randomly sets in upon formation of the domain. In general, the form of the order parameter is dependent on the symmetries of the system in question. It may be a complex scalar quantity (e.g.,~in superfluid $^4$He), a vector (e.g.,~spin or liquid crystal systems), or can have a more complex tensor form (e.g.,~superfluid $^3$He). In all cases, the order parameter is continuous within a domain, but with singular points along the boundaries between domains. Such discontinuities in the order parameter field are called defects. A defect is said to be topological if removing it requires not merely local perturbation, but complete destruction of the macroscopic order \cite{Mermin1979}. In 3D systems, topological defects may take the form of lines \cite{Mermin1979}, and arise during a quench if the intersection of three or more domains has the correct structure. In the case of a superfluid, line defects form when the complex phase winds by $2\pi$ around the line. If a line of intersection between three or more domains is such that the phase increases sequentially around the center, with a total change in phase close to $2\pi$ (see Figure \ref{fig:Domain_Cartoon}), then it may nucleate a vortex once the system reaches equilibrium.

In its simplest form the KZ theory predicts that the number of defects that form will be set by the average domain size at the freeze-out time $\hat{\xi}$. The areal density of vortices predicted by this theory is $L=\hat{\xi}^{-2}$, based on the number of domains that are expected to form across a given surface area. This prediction has been studied in a number of systems such as liquid crystals \cite{Chuang1991}, Bose-Einstein condensates \cite{Sadler2006,Scherer2007,Lamporesi2013,Anquez2016,goo2021}, strongly interacting Fermi gasses \cite{ko2019}, colloidal monolayers \cite{Deutschlander2015}, and quantum annealing devices \cite{bando2020}.

Surprisingly, despite superfluid $^4$He being the original model system for Zurek's theory, experiments in $^4$He have not been compatible with the theory \cite{Hendry1994,Dodd1998}. The observed density of vortices was at least two orders of magnitude smaller than predicted \cite{Dodd1998}. The reason for this discrepancy is currently not well understood. More success has been found in studying superfluid $^3$He. The order parameter of $^3$He has a more complex form, including not just a complex phase, but a matrix structure related to the angular momentum degrees of freedom of Cooper pairs \cite{Vollhardt13}. This may result in multiple types of vortices \cite{salomaa1987}, as well as domain wall structures \cite{salomaa1988,walmsley2004,gluscevich2024kibble}. The types of defects that form depend on the superfluid phase in question (i.e., $^3$He-A, $^3$He-B, etc.), as well as the boundary conditions of the system \cite{SI}. The KZM has been studied in $^3$He using nuclear magnetic resonance to detect vortices \cite{Ruutu1996,Bauerle1996,Autti2016,Autti2020,rysti2021suppressing}.

\begin{figure}
    \centering
    \includegraphics[width=0.9\linewidth]{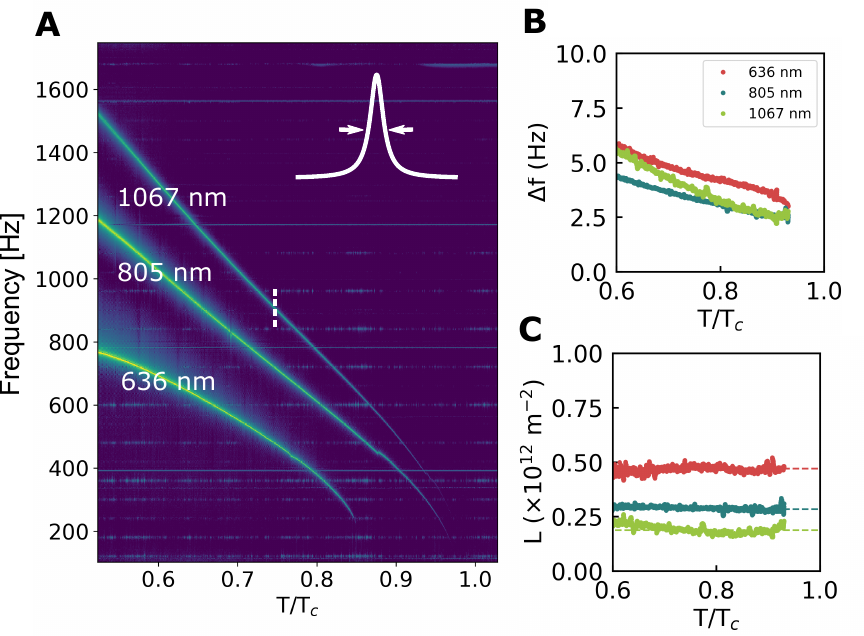}
    \caption{\textbf{Helmholtz Resonator Data.} \textbf{(a)} Resonance curves of the three different devices, with varying nanoscale confinements, measured during a temperature sweep carried out at 15 bar. Inset shows the Lorentzian resonance shape at a single temperature. From the center frequency, the superfluid density can be extracted \cite{Shook2020}.
    \textbf{(b)} Additionally, the linewidth of each device can be extracted, plotted here as a function temperature. \textbf{(c)} Resulting vortex densities calibrated from the Helmholtz resonator linewidths and the mutual friction, see main text. Note the lack of temperature dependence, and the dependence on confinement.}  
    \label{fig:Linewidth_Data}
\end{figure}

Here, we build on previous research in quasi-two-dimensional Bose gases \cite{Chomaz2015emergence,Chomaz2016exploring} by studying the formation of defects due to a phase transition in $^3$He that has been confined to a thin channel, where the smallest dimension is less than the KZM lengthscale. This is a regime where the finite size of the geometry may modify the formation of defects in the complex order parameter of $^3$He. 

\section*{Methods}

We have designed an experiment sensitive to vortex dissipation that is distinct from previous studies in helium. It consists of nanofluidic devices, known as Helmholtz resonators, which have been described in detail elsewhere \cite{Rojas2015,Souris2017,Shook2020}. These devices allow for confinement of the helium in channels of thickness $H_{1,2,3} =$ 636, 805, and 1067 nm,
respectively. Additional follow-up experiments were later performed with a $750$ nm device. Parallel-plate capacitors patterned onto the devices allow for an electrostatic force, which slightly compresses the center of the device, modifying the volume, and creating a pressure gradient down the channel. This drives superfluid motion in the channels, but notably does not couple to normal-fluid motion due to the viscous penetration depth being large compared to the thickness of the channel \cite{SI}. By driving this superfluid Helmholtz mode, we are able to extract two relevant parameters: a center frequency of the Helmholtz mode, $f_0$, the square of which is proportional to the superfluid density, $\rho_s \propto f_0^2$ (i.e., the mass density of the superfluid condensate, excluding the normal fluid contribution); and a linewidth, $\Delta f$, which characterizes the resonator's dissipation. Previously, phase transitions have been observed as kinks in the Helmholtz resonance frequency, allowing mapping of the phase diagram, as described in Reference~\cite{Shook2020}.

Here, we focus on the dissipation in these devices and find that they exhibit linewidths of several Hz (see Figure 2), corresponding to quality factors of a few hundred depending on the temperature and geometry in question. The quality factor is many orders of magnitude lower than that measured in similar devices in $^4$He \cite{Rojas2015}. For this reason, the dissipation must be limited by a process intrinsic to the $^3$He system, rather than the quartz substrate that the Helmholtz resonators are fabricated from (which should be much lower in the millikelvin temperature regime compared to the 1 K regime $^4$He experiment \cite{Rojas2015}). We have identified possible dissipation mechanisms which are intrinsic to the superfluid phase of $^3$He: surface-bound state dissipation, vortex mutual friction, normal fluid slip, and second viscosity. Additionally, there are phase-specific effects associated with $^3$He textures. In a prior publication \cite{shook2024surface}, we have investigated  critical velocity by studying the resonance amplitude's drive dependence. A linear regime exists where the amplitude is proportional to the drive, and a change in the slope of the amplitude-force curve marks a non-linear regime. All of the measurements in this publication were performed in the linear regime for both the A and B-phases where we do not expect dissipation due to surface bound-state dissipation. Calculations regarding normal fluid slip, second viscosity, and textural effects have been presented in the Supplementary Material \cite{SI}. We find that these effects are too small to explain the observed linewidth. Furthermore, these effects do not exhibit the observed temperature dependence of the linewidth. 

The best explanation for the observed dissipation in the linear regime is vortex mutual friction. The basis for this is that the temperature dependence of the ratio, $\Delta f/f_0^2$, is well fit by a function linear in the quantity $\alpha(T)$
\begin{equation}
    \frac{\Delta f(T)}{f_0^2(T)} = C_1 \alpha(T) + C_2.
    \label{eq:offset_model}
\end{equation}
Here, $\alpha$ is the mutual friction parameter associated with dissipative vortex motion, which is calculated following Ref.~\cite{Bevan1997} using the theoretical model developed by \cite{kopnin1995spectral,Kopnin1995}. The basic idea is that when the superfluid flows, it induces motion in the vortex lines due to a Magnus force. When a vortex line moves relative to the normal fluid component (which here is static), dissipation is generated due to scattering of excitations with the vortex core. Further details can be found in the Supplementary Material \cite{SI}. Note that the equation used to calculate $\alpha$ depends on an empirical proportionality factor measured by Bevan et al.~\cite{Bevan1997}. Differences in our experiment could affect the magnitude of $\alpha$, but it will remain within the same order of magnitude.
When the experimentally measured ratio $\Delta f(T)/f_0^2(T)$ is plotted against the theoretical $\alpha(T)$, a clear linear trend is observed. To explain this trend, a mutual friction model for our device is introduced, as outlined in the Supplementary Material \cite{SI}. This model predicts that the slope of Equation~\ref{eq:offset_model}, $C_1$, should be linear in $L$, with a temperature-independent proportionality factor $\gamma$ that is determined by geometric and material parameters. Once this parameter is known, the vortex density can be computed from either fitting constant as $L=C_1/\gamma = (\Delta f/f_0^2 - C_2)/\gamma \alpha$.

\begin{figure}
    \centering
    \includegraphics[width=0.9\linewidth]{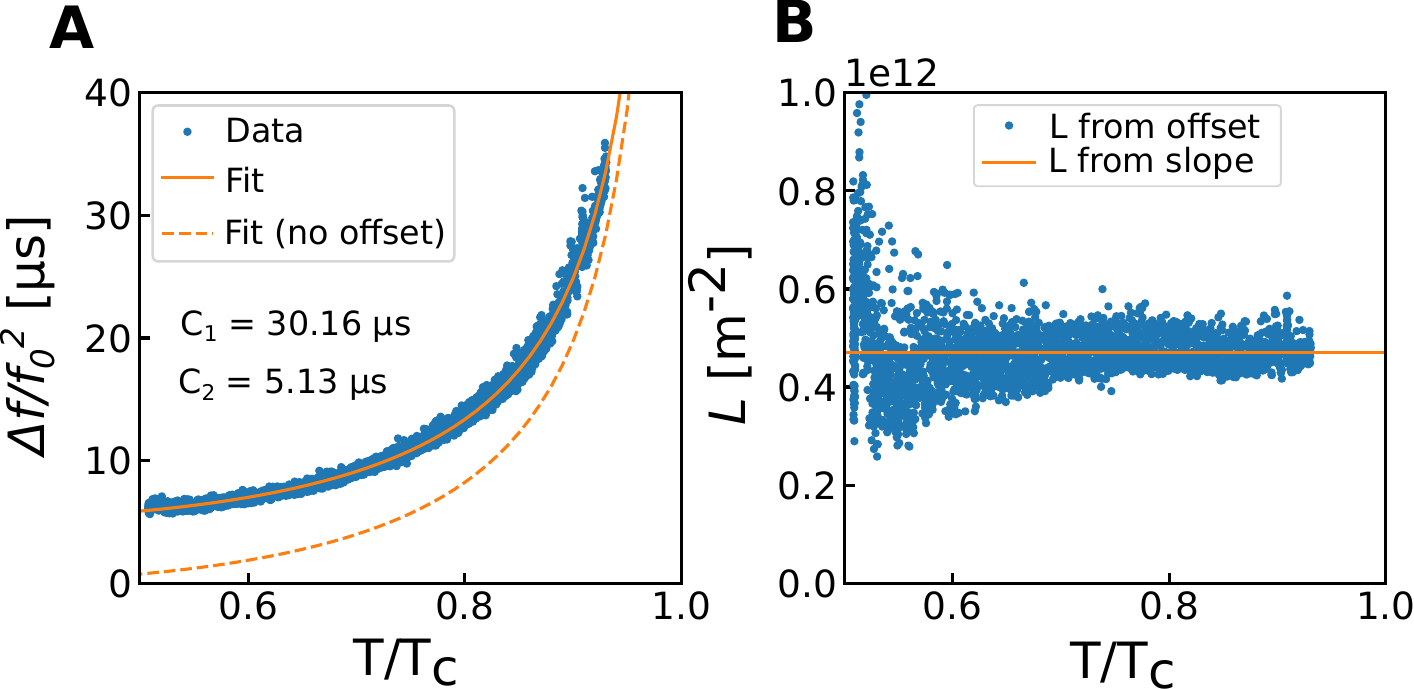}
    \caption{\textbf{Vortex Density Calibration Fit.} \textbf{(a)} Plot of the experimental ratio $\Delta f/f_0^2$ as a function of temperature at 15 bar for the 636 nm device. This data has been fit to our model and plotted as a solid orange line. The dashed line displays the same curve if the offset is omitted in order to highlight the modification. \textbf{(b)} The vortex density is computed from the fit parameters $C_1$ and $C_2$.}
    \label{fig:alpha_fit}
\end{figure}

An example of the fit to the data can be seen in Figure \ref{fig:alpha_fit}. The constant offset $C_2$ is interpreted as being due to other loss mechanisms that scale proportionally with the superfluid density $\rho_s \propto f_0^2$. There are a number of candidates for this, which are discussed in the Supplementary Material \cite{SI}.

\section*{Discussion}

\begin{figure}
    \centering
    \includegraphics[width=0.8\linewidth]{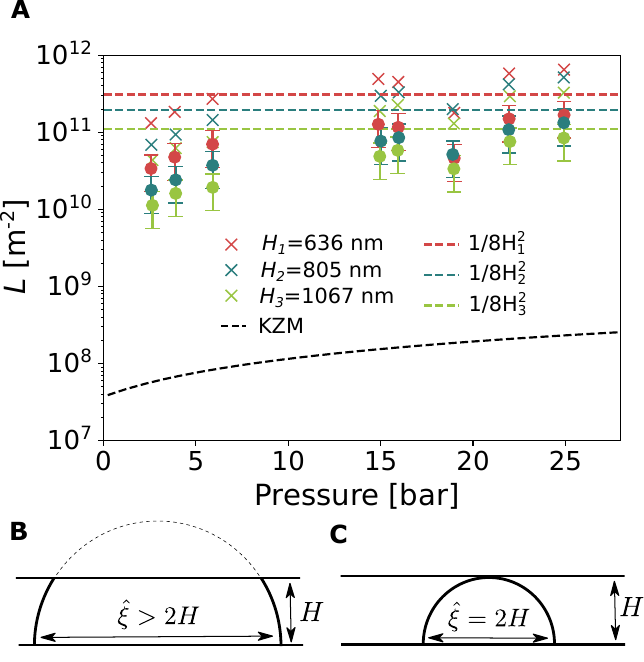}
    \caption{\textbf{Comparison of the KZ theory and truncation model.} \textbf{(a)} The prediction of the KZ theory, for a ramp rate of $dT/dt = 0.12$ mK/hr, is plotted as a dashed black curve as a function of pressure. The superfluid coherence length, and therefore also $\hat{\xi}$, changes with pressure. This curve is compared to the experimental vortex densities --- shown as a function of pressure for 636 nm (red), 805 nm (blue), and 1067 nm (green) devices --- and found to differ by more than three orders of magnitude.  The truncated length-scale prediction of $L$ is shown as colored dashed lines. The x markers represent the fits to the raw data, and the circles are after correction for the artificial broadening of the Helmholtz resonances (see the discussion in the supplementary material \cite{SI}). \textbf{(b)} Cartoon diagram of a domain that propagates from an initial point to a distance $\hat{\xi} > 2H$ (diameter), The walls of the domain become plane-like due to the confining walls. \textbf{(c)} A domain which has propagated a distance $\hat{\xi} = 2H$. This is when vortex lines joining the two walls first become possible.}
    \label{fig:Comparing_Models}
\end{figure}

Using this mutual friction model, a vortex density is extracted from the fit parameters of the Helmholtz resonance. The result is then compared to the predicted defect density of the KZ theory. It is worth pointing out that the confined dimension of the devices ($H = 636-1067$ nm) is orders of magnitude smaller than the KZ length scale  of $\hat{\xi} = 60-160$ $\mu$m. Because of this, it is expected that during the phase transition the propagating domains should rapidly reach both walls. At this point the domains cannot propagate further in the confined direction but should continue propagating in the two macroscopic directions. In this case the density of defects is determined by the KZ theory (see the Supplementary Material \cite{SI}) to be $L = \hat{\xi}_{\textrm{KZM}}^{-2} = \xi_0^{-2}\sqrt{\tau_0/\tau_Q}$
where $\xi_0$ is the zero temperature coherence length, $\tau_0$ is the relaxation time and $\tau_Q = T_c(dT/dt)^{-1}$ is a time constant characterizing the temperature ramp rate $dT/dt$. 
For the temperature ramp rate used in the majority of the experiments, 0.12 mK/hour, this corresponds to $L \sim 10^{7} - 10^{8}$ m$^{-2}$ over the range of pressures investigated in this experiment. Contrary to this, vortex densities are measured on the order of $L \sim 10^{10} - 10^{11}$ m$^{-2}$, as shown in Figure \ref{fig:Comparing_Models}. Also notable is the clear dependence of vortex density on the confinement of the device; with the most confined channel ($636$ nm) having the highest density and the least confined ($1067$ nm) the lowest. Geometric effects do not enter into the original KZ theory, which should depend only on the temperature ramp rate and the pressure. Such high vortex densities (and therefore small domain sizes) would require extraordinarily fast quench rates (approximately 50-200 mK/s depending on the device) if they were dictated by the freeze-out time of the KZ theory.

Although the calibrated vortex area density does not show agreement with the associated KZ length scale, there is a close correspondence with the height of the channel. If the KZ length scale is replaced with an effective ``truncation" length $\hat{\xi}_{\textrm{KZM}} \to 2H$, and the vortex area density is estimated in an analogous manner as $L=H/\hat{\xi}^3 = (8H^{2})^{-1}$, then a vortex density results that is within an order of magnitude of the calibrated value, Fig.~4. The choice of $2H$ is motivated by the observation that this is the spatial extent of a domain in the non-confined dimension once the domain has had time to propagate from one wall to the other. This observation is suggestive and seems to imply a qualitatively different process of defect formation than the KZ picture. Although we can only speculate at the non-equilibrium process by which the domains evolve near the transition, note that, unlike the bulk system, the confined channels are expected to produce vortices that terminate at the walls, Fig.~4. At equilibrium, after the phase transition, this should stabilize vortex lines since, unlike a loop, a wall-terminated line cannot shrink indefinitely to the point where it annihilates itself.

Before the freeze-out time, one cannot meaningfully speak of topological defects, since the ordered equilibrium phase has not been established; however, one can speak of statistically independent domains with ``proto-defects" at the vertices of domains. It is speculated that the presence of the walls may stabilize proto-defects during the period of time after the domains are large enough to bridge the two walls, but before the domains have had time to spread and merge in the in-plane directions. This leads to an ensemble of vortices where the mean separation between cores is determined by the size of the channels, rather than the freeze-out time. It is also possible that surface roughness plays a role in stabilizing the formation of defects. Deeper theoretical investigation is required to conclusively resolve this question.

Some degree of overestimation is expected as in the original KZ theory, since not every vertex between domains will nucleate a vortex. This is evidenced in numerical investigations, which have found that the KZ prediction overestimates the defect density and that the true value is obtained by scaling $\hat{\xi}$ by a factor of 5-10 \cite{das2012winding,laguna1997density,yates1998vortex,antunes1999vortex,bettencourt2000ginzburg}. Such scaling has also been observed in $^3$He NMR experiments where it is estimated $\hat{\xi}$ should be scaled by 2.3 times the KZM value \cite{rams2019symmetry,rysti2021suppressing}. 
Finally, note that some vortices may annihilate at the boundaries, or due to vortex recombination during the phase transition \cite{eltsov2010vortex}. The latter mechanism is expected to depend on the coherence length, which may explain the pressure dependence in the vortex density we observe \cite{SI}.

Unlike the KZ model, where the vortex density is determined by a freeze-out time, the interpretation here depends only on a spatial length-scale and is therefore ramp rate independent. After initial measurements of the 636, 805, and 1067 nm devices, additional follow-up experiments were performed using the 750 nm device to investigate any possible ramp rate dependence. This was achieved by varying the temperature ramp rate from 0.03-0.36 mK/hr while passing through the critical temperature. The maximum temperature ramp rates possible were limited both by the thermal time constant of the experimental cell, and the requirements that the nuclear demagnetization process used to cool the apparatus be adiabatic. Assuming the KZ scaling of $L \propto \sqrt{dT/dt}$, a factor of 10 increase in the ramp rate will increase the vortex density by $\sim$3 times, which should have been observable in the experiments. Despite this, no evidence of a change in linewidth is seen when the temperature was ramped at different rates through $T_c$ \cite{SI}.


Ramp rate independent remanent defects have also been observed in a Bose-Einstein condensate system with a quasi-2D trapping potential \cite{Chomaz2015emergence,Chomaz2016exploring}. For fast ramp rates, they observed a scaling compatible with the KZ mechanism, but at low ramp rates, they found that the defect density saturates when the KZ domain size would be comparable with the smallest dimension of their system. The authors speculate that for slow cooling, the small scale variations in the trapping potential might dictate the probability of forming remanent vortices. This is not dissimilar to the present experiment, except that here the temperature ramp is in a regime where the KZ length scale is always large compared the confined dimension. It would be interesting in future experiments to investigate the role of surface roughness on vortex formation using an established technique of pre-plating the surfaces with $^4$He \cite{Freeman}. Also potentially relevant are the observations of Xia et al.~\cite{xia2020winding} in a superconducting ring, where below a critical circumference the defect scaling becomes ramp rate independent. It should be noted that these observations are distinct from truly 2D systems which lack a transverse length scale, such as \cite{Deutschlander2015}, and may still exhibit KZ scaling with a different dimension.

\section*{Conclusion}

We have performed measurements of the dissipation of $^3$He fourth-sound resonances, which is consistent with mutual friction due to a fixed density of remanent vortices. The density of vortices is found to be independent of temperature ramp rate, but is a function of the channel thickness. A model of vortex formation, distinct from the KZ model, is proposed in which a spatial length scale of the system sets the estimated defect density rather than the quench rate as is the case in the KZ model. This interpretation is found to be within an order of magnitude of the observations, comparable to the agreement seen in tests of the KZM, and the scaling with confinement is in reasonable agreement. These measurements are part of a growing body of work demonstrating that remanent defect formation is modified in systems where one or more dimensions are small compared to the KZ length scale \cite{rysti2021suppressing,Chomaz2015emergence,xia2020winding} and motivate further studies into phase transitions in non-infinite systems.


\begin{acknowledgments}
The authors acknowledge that the land on which this work was performed is in Treaty Six Territory, the traditional territories of many First Nations, Métis, and Inuit in Alberta. They acknowledge the support from the Canada Foundation for Innovation; the Natural Sciences and Engineering Research Council, Canada (Grant Nos.~RGPIN-2022-03078,  ALLRP 602120-2024); and Alberta Innovates (Grant No.~242506367).
\end{acknowledgments}


\end{document}


\preprint{APS/123-QED}

\title{Supplementary Material}

\author{Alexander J. Shook}
\email{ashook@ualberta.ca}
\author{Daksh Malhotra}%
\author{Aymar Muhikira}%
\author{Vaisakh Vadakkumbatt}%
\author{John P. Davis}%
\email{jdavis@ualberta.ca}
\affiliation{%
 Department of Physics, University of Alberta, Edmonton, Alberta T6G 2E1, Canada}%


\maketitle

\section{$^3$He Defects}

Superfluids are universally characterized by long-range phase coherence associated with broken $U(1)$ gauge symmetry. In a superfluid, such as $^4$He where this is the only type of coherence present, the order parameter is proportional to a complex phase factor $e^{i\phi(\vec{r})}$ and the superfluid hosts a defect whenever $\phi(\vec{r})$ winds about a point by a multiple of $2\pi$. The order parameter of $^3$He is significantly more complicated due to the constituent Cooper pairs having spin-triplet, $p$-wave pairing (i.e., $S=L=1$). Each Cooper pair therefore has spin and angular momentum vectors which may also exhibit spatial coherence. This leads to a number of distinct superfluid phases, as well as numerous possible defects. 

The full symmetry group of the normal fluid phase is 
\begin{equation}
    G = SO_L(3) \times SO_S(3) \times U_{\phi}(1),
\end{equation}
where $SO_L(3)$ is the symmetry group associated with rotations of the orbital angular momentum, $SO_S(3)$ is associated with the spin, and $U_{\phi}(1)$ with the phase. In the superfluid B-phase, the spin-orbit interaction internal to the Cooper pair requires a fixed angle between the spin and orbital angular momentum degrees of freedom. The system, therefore, retains an $SO(3)$ invariance but under joint rotation of both degrees of freedom
\begin{equation}
    H_B = SO_{L+S}(3) \times U_{\phi}(1).
\end{equation}
In the A-phase, the invariance under rotations in spin and orbital spaces is broken separately. Given an orientation of the orbital anisotropy vector $\vec{\ell}$, the spin anisotropy vector $\vec{d}$, may form an arbitrary angle with $\hat{\ell}$ in the plane containing both vectors. Rotations of this plane around $\hat{\ell}$ are related to the superfluid phase, so the residual symmetry of the phase can be shown to be 
\begin{equation}
    H_A = \mathbb{Z}_{2} \times U(1) \times U(1).
\end{equation}
Here, the group $\mathbb{Z}_{2}$ comes from the fact that the spin-orbit coupling scales proportionally to $(\hat{d}\cdot\hat{\ell})^2$, such that there is an indifference to the sign. In the case of a zero magnetic field, the two vectors become locked, $\hat{d} \parallel \pm \hat{\ell}$, due to the spin-orbit interaction. This breaks one of the $U(1)$ symmetry groups.

A complete list of defects that may form comes from an analysis of the symmetry group of a particular phase. We are chiefly interested in the cases of the A-phase and the (planar distorted) B-phase relevant to our experiment. For the A-phase the relevant degrees of freedom are the orbital and spin anisotropy vectors, $\hat{\ell}$ and $\hat{d}$, in addition to the phase $\phi$. In the B-phase the degrees of freedom are the relative spin-orbit angle $\theta$, rotation axis $\hat{n}$, and phase $\phi$. The defects in these fields may be either plane-like, line-like, or point-like \cite{Vollhardt2013}.

\subsection{A-Phase Line Defects}
\label{sec:3He-A_Defects}

$^3$He-A hosts three topological classes of line defects, characterized by a topological charge $N= 0, \pm 1/2, 1$ \cite{Salomaa1987}, corresponding to textures, half-quantum vortices, and singular vortices, respectively. Formally, one can express these defects in terms of the $^3$He-A order parameter, which is a $3\times 3$ complex matrix of the form
\begin{equation}
    A_{\mu j} = \Delta \hat{d}_{\mu} (\hat{e}_{1,j}+i\hat{e}_{2,j})e^{i\phi},
\end{equation}
where $\Delta$ is the superfluid gap, $\hat{d}$ is the direction of spin anisotropy for Cooper pairs, and $\phi$ is the phase. The unit vectors $\hat{e}_{1}$ and $\hat{e}_{2}$, are perpendicular vectors that define an ``orbital triad" such that the direction of the orbital angular momentum quantization axis is given by $\hat{\ell} = \hat{e}_{1} \times \hat{e}_{2}$. The A-phase is characterized by having long-range correlation in both the $\hat{\ell}$ and $\hat{d}$ vector fields.

\begin{figure}
    \centering
    \includegraphics[width=\linewidth]{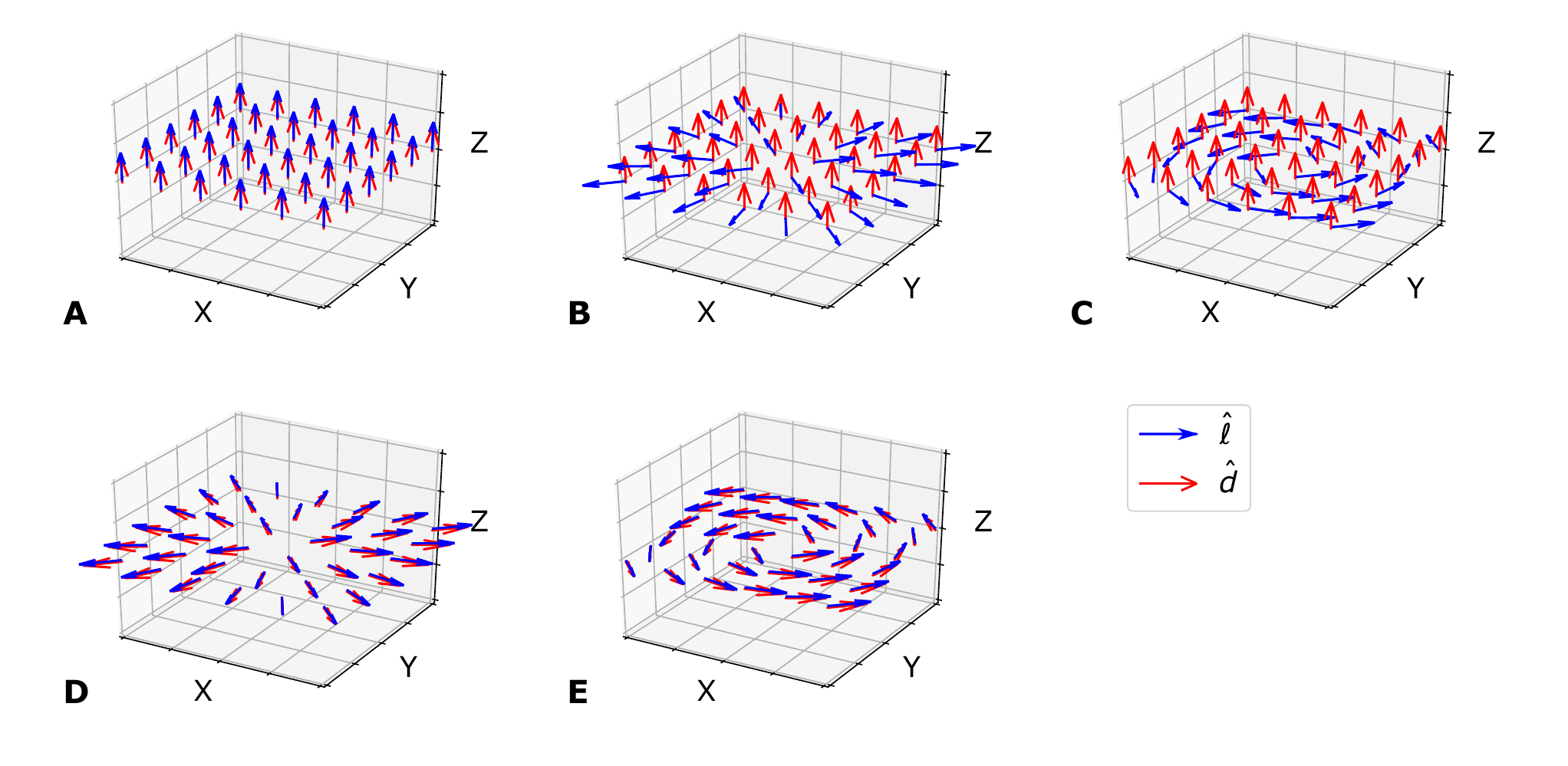}
    \caption{\textbf{A phase N=1 defects.} Cross-sections of the five distinct defects with topological charge $N=1$ that occur in $^3$He-A. The orbital angular momentum quantization axes $\hat{\ell}$ are plotted in blue with narrow arrowheads, and the spin quantization axes $\hat{d}$ are plotted in red with wide arrowheads. The pure phase vortex, depicted in \textbf{(A)}, exhibits uniform spin and orbital vector fields. In other defects, the $\hat{\ell}$ field may be radial (see \textbf{(B)},\textbf{(D)}) or azimuthal (see \textbf{(C)},\textbf{(E)}), with the $\hat{d}$ field either parallel (see \textbf{(D)},\textbf{(E)}) or perpendicular in the $\hat{z}$ direction (see \textbf{(B)},\textbf{(C)}).}
    \label{fig:3He-A_Defects}
\end{figure}

Boundary conditions play a large role in determining the energy of the texture that forms. At a wall, the orbital angular momentum vector $\vec{\ell}$ is constrained to be normal to the surface \cite{Vollhardt2013}. Further from the wall $\hat{\ell}$ may rotate, however gradients in the vector field impose an energy cost. A comparison of the gradient energy to bulk orientational effects sets a healing length $\xi^A_{\scriptsize\textrm{heal}}$, which is the distance from the wall at which the superfluid becomes bulk-like. In $^3$He-A this length scale is estimated to be $\xi^A_{\scriptsize\textrm{heal}} \sim 8$ $\mu$m \cite{Leggett1975}. In a parallel plate geometry where the separation between the walls is small compared to $\xi^A_{\scriptsize\textrm{heal}}$ a uniform $\vec{\ell}$ field is strongly preferred energetically. In our experiment, the largest plate separation is $1$ $\mu$m which is $1-2$ orders of magnitude smaller than the healing length, hence strongly favoring an $\vec{\ell}$ orientation as shown in Fig.~\ref{fig:3He-A_Defects}A.

Although non-uniform textures are strongly disfavored from an energetic perspective, non-trivial textures can emerge from the phase transition process. Once the system is at equilibrium, the texture is expected to relax into the uniform state if possible; however, for topologically non-trivial textures, this is impossible. This will result in defects in the order parameter that persist. For $^3$He-A, textures with $N=0$ are topologically equivalent to the uniform texture, and therefore can likely be ruled out on energetic grounds.

 We then turn our attention to $N=1$ defects, of which there are five. Expressed in a cylindrical coordinate system, $(r,\theta,z)$, centered on the vortex core, these defects can be expressed as
\begin{equation}
    \vec{d} = \vec{\ell} = \hat{z}, \quad \hat{e}_{1}+i\hat{e}_{2} = \hat{x}+i\hat{y}, \quad \phi = (2n-1)\theta,
\end{equation}
\begin{equation}
    \vec{d} = \vec{z}, \quad \vec{\ell} = \hat{r}, \quad \hat{e}_{1}+i\hat{e}_2 = \hat{z}-i\hat{\theta}, \quad \phi = 2n\theta,
\end{equation}
\begin{equation}
    \vec{d} = \vec{z}, \quad \vec{\ell} = \hat{\theta}, \quad \hat{e}_1+i\hat{e}_2 = \hat{z}+i\hat{r}, \quad \phi = 2n\theta,
\end{equation}
\begin{equation}
    \vec{d} = \vec{\ell} = \hat{r}, \quad \hat{e}_1+i\hat{e}_2 = \hat{z}-i\hat{\theta}, \quad \phi = 2n\theta,
\end{equation}
\begin{equation}
    \vec{d} = \vec{\ell} = \hat{\theta}, \quad \hat{e}_1+i\hat{e}_2 = \hat{z}+i\hat{r}, \quad \phi = 2n\theta,
\end{equation}
where $n$ is an integer \cite{Salomaa1987}. These defects are depicted in Fig.~\ref{fig:3He-A_Defects}. Of the five defects, one can see that all but the first involve $\hat{\ell}$ fields that are either radial or azimuthal. In order for these defects with orbital dysgyrations to comply with the boundary conditions at the wall, $\hat{\ell}$ must change approaching the surface. Although one can orient the vortex core in a number of ways relative to the surface, all clearly require tight bending of $\hat{\ell}$ across the slab, which is many times smaller than the healing length. For this reason, we expect there to be a high energy cost associated with the formation of vortices that involve bending of $\hat{\ell}$ texture. This does not necessarily prohibit the formation of such defects during phase transition, and such defects are expected to be topologically stable.

Despite this, there is reason to think that the formation of vortices where $\hat{\ell}$ is a function of $z$ is suppressed. Here we can make an analogy to the formation of Bose-Einstein condensates (BEC) in strongly anisotropic confining fields. In such systems the critical temperature at which a condensate forms for momentum states with a finite momentum in the highly confined dimension is different than that for momentum states with zero projection in this direction \cite{Druten1997,Chomaz2015}. For very highly confined systems (or ones with a smaller number of particles) only the states in the plane perpendicular to the confined dimension condense. In the $^3$He system we expect an analogous situation to occur. Cooper pair may condense into different angular momentum states. Near the surfaces of the system, Cooper pairs may only condense into a state with orbital angular momentum perpendicular to the surface. For points in the middle of the slab this argument does not hold, however, because the KZM domain size $\hat{\xi} \gg H$, correlation between the surface states and those in the middle of the slab is rapidly established. For this reason, we do not expect any defects to form which require gradients of $\hat{\ell}$ in $z$. Using the language of \cite{rysti2021suppressing}, the role of confinement can be viewed as a symmetry-violating bias, which suppresses the formation of these types of defects. 

Here we also note that if other varieties of $N=1$ vortices were to form, we would expect them to be static. This is because the energy associated with such tight bending of the $\hat{\ell}$ is much larger than the energy associated with the flow velocity.
In order for these two energies to be comparable, the flow velocity needs to be at least as large as the critical velocity predicted by de Gennes and Rainer \cite{de1974alignment}
\begin{equation}
    v_{\textrm{Fr}} = \frac{3}{4} \frac{\pi\hbar}{2m^{*}H} \approx 27 - 45 \textrm{ mm/s}.
\end{equation}
In fact this number likely significantly underestimates the velocity threshold since it assumes a gradient in only one dimension, but the line defects require gradients in all three dimensions. In a previous publication \cite{shook2024surface} we show that the A-phase linear regime exists at much lower velocities, so if vortices with orbital gradients exist, they are are likely ``frozen-in" by the confinement. If the defects are static, then they do not contribute to the mutual friction and, thus, the total dissipation.

Lastly, defects with $N = \pm 1/2$, known as half-quantum vortices, may form in $^3$He-A in a parallel plate geometry with a thickness smaller than the dipole length $\xi^A_D \sim 8$ $\mu$m, which is realized in our experiment. The form of the vortex is specified by 
\begin{equation}
    \hat{e}_1+i\hat{e}_2 = e^{i\gamma}(\hat{x}+i\hat{y}), \quad \gamma = 2k \pm \frac{1}{2},
\end{equation}
\begin{equation}
    \hat{d} = \cos(\gamma'\theta)\hat{x} + \sin(\gamma'\theta)\hat{y}, \quad \gamma' = k'+\frac{1}{2},
\end{equation}
where $k$ and $k'$ are integers \cite{Vollhardt2013}. These half-quantum vortices must form in pairs, separated by a characteristic distance $r_{1/2}$, as the terminating points of two domain walls. These walls separate regions with opposite orientations of $\hat{d}$. When generated by the Kibble-Zurek mechanism, the separation between half-quantum vortices is expected to be $r_{1/2} = \hat{\xi}$  \cite{rysti2021suppressing}.

Experiments seeking to generate half-quantum vortices via rotation in a parallel plate geometry also require a magnetic field parallel to the surface normal, where the field strength is at least $|\vec{H}| \sim 30$ G \cite{Vollhardt2013,Autti2016}. This orients $\hat{d}$ vectors perpendicular to the magnetic field and surface normal. In the absence of such a field, the orientation of $\hat{d}$ will be determined by spin-orbit interaction, which favors a uniform orientation parallel to $\hat{\ell}$ and the surface normal. As with the $N=1$ vortices, this energetic argument is not sufficient to rule out the formation of half-quantum vortices during the phase transition, since it is a fundamentally non-equilibrium process. Nevertheless, the spin-orbit interaction internal to each Cooper pair exists even before thermodynamic equilibrium is established. This reduces the total symmetry group of the A-phase to $\mathbb{Z}_{2} \times U(1)$, and means that within each coherent domain that forms, we can expect $\hat{d} \parallel \pm \hat{\ell}$. This should lead to the formation of domain walls with opposite signs for $\vec{\ell}$ and/or $\hat{d}$, however, the domain structure is two-dimensional \cite{gluscevich2024kibble}. This is not compatible with half-quantum vortices, which require $\hat{d}$ to lie in the $xy$-plane. We therefore argue that the formation of half-quantum vortices should be suppressed for the same reason as $N=1$ vortices with gradients in $\hat{\ell}$. Domains forming at the surface will only allow for states where both vectors are normal to the surface, and coherence will rapidly be established between the two walls since $\hat{\xi} \gg H$. We note that this claim is consistent with the assumptions of \cite{gluscevich2024kibble}.


\subsection{A-phase Planar Defects}

For an A-phase sample confined between parallel plates, the boundary conditions require $\hat{\ell}$ to be normal to the walls. Both states $\pm \hat{\ell}$ (i.e., pointing into or out of the wall) are allowed, however, and there is no energetic difference between them. This means that during phase transition into the A-phase, one should expect regions of both signs to form throughout the sample, resulting in domain walls \cite{halsey1985topological,walmsley2004intrinsic}. The existence of a domain wall increases the energy of the system, due to the $\hat{\ell}$ gradient between the domains, but it is protected by its topology such that it is stable. As with the orbital gradient vortices, however, we do not expect A-phase domain walls to be dynamic due to the largeness of the gradient energy associated with the domain wall and therefore do not expect them to contribute to the dissipation.

\subsection{B-phase Line Defects}

In the B-phase there are three types of line-like defects; two of which are believed to be thermodynamically stable \cite{mineev1986superfluid}. The first is a pure phase vortex with a normal fluid core (i.e. the superfluid gap closes in the core), referred to as the o-vortex in the literature. In the second case, the gap does not close in the vortex core, but rather the order parameter takes on the form of the A-phase. These are known as a v-vortices or A-core vortices. Lastly, there is a double-core vortex where the gap is suppressed at two points and has a polar phase structure in each core, which is called D-core. In addition to these line-like defects, there is also a soliton structure that has been observed in NMR experiments \cite{eltsov2000composite}.

Ginzburg-Landau calculations show that the o-vortex is in fact never thermodynamically stable. It has been demonstrated both experimentally \cite{hakonen1983magnetic}, and theoretically \cite{volovik1985spontaneous,salomaa1986vortices}, that there is a region of the phase diagram at high pressures, and near the $T_{AB}$ line where the A-core vortices are metastable, but at low temperatures ($<$ 1.4-1.6 mK) the vortices transition to D-core. At lower pressures vortices are always D-core.

The experiment by Beven et al. \cite{Bevan1997} has shown a small change in dissipation due to this vortex core transition, though the effect was initially too weak to see until special modifications to the experiment were made. We do not see such a signature in our data sets, likely due to a lack of measurement sensitivity. For this reason, we do not distinguish between A-core and D-core vortices in our mutual friction model since the difference in dissipation seems to be too small to observe.

\section{Mutual Friction}

Here we outline a theory of vortex mutual friction based on the Hall-Vinen-Iordanskii theory of vortex motion \cite{Hall1970,Vinen2001}. If we consider a rectilinear vortex filament of length $H$ in a superfluid flowing perpendicular to the vortex axis of rotation, the vortex feels two forces. The circulation of the vortex adds with the external velocity field on one side of the vortex, and subtracts on the other, resulting in a pressure differential which pushes the vortex. This is the Magnus force $\vec{F}_M$, which is perpendicular to the relative velocity of the vortex core $\vec{v}_L$ to the superfluid flow velocity $\vec{v}_s$:
\begin{equation}
    \frac{\vec{F}^{(M)}}{H} = \kappa \rho_s (\vec{v}_s-\vec{v}_L) \times \hat{z}.
    \label{eq:MagnusForce}
\end{equation}
Here $\rho_s$ is the superfluid density. The second force, $\vec{F}_N$, is due to the normal fluid, and arises from interaction of quasiparticles with vortex bound states,
\begin{equation}
    \frac{\vec{F}^{(N)}}{H} = D(\vec{v}_n-\vec{v}_L) + D' \hat{z}\times(\vec{v}_n-\vec{v}_L).
\end{equation}
The constants $D$ and $D'$ represent the magnitudes of the dissipative and reactive components of this force respectively. 

The vortex line carries negligible inertia so in the absence of additional forces, the Magnus force and normal fluid forces sum to zero
\begin{equation}
    \vec{F}^{(M)} + \vec{F}^{(N)} = 0.
\end{equation}
Re-arranging this equation gives an expression for the velocity of the vortex line 
\begin{equation}
    \vec{v}_L = \vec{v}_s + \alpha \hat{z} \times (\vec{v}_n - \vec{v}_s) + \alpha'(\vec{v}_n - \vec{v}_s).
    \label{eq:vL}
\end{equation}
Two mutual friction parameters have been defined
\begin{equation}
    \alpha = \frac{d_{\parallel}}{d_{\parallel}^2+(1-d_{\perp})^2}, \qquad 1-\alpha' = \frac{1-d_{\perp}}{d_{\parallel}^2+(1-d_{\perp})^2},
\end{equation}
where
\begin{equation}
    d_{\parallel} = \frac{D}{\kappa \rho_s}, \qquad d_{\perp} = \frac{D'}{\kappa \rho_s}.
\end{equation}
Substituting equation \ref{eq:vL} into equation \ref{eq:MagnusForce} gives a Magnus force per unit length of
\begin{equation}
    \frac{\vec{F}^{(M)}}{H} = \kappa \rho_s \alpha (\vec{v}_s-\vec{v}_n) - \kappa \rho_s \alpha' \hat{z} \times (\vec{v}_s-\vec{v}_n).
\end{equation}
Now suppose that $N_L$ such vortices exist in a rectangular channel of height $H$, width $w$, and length $\ell$, such that the channel volume is $V_c = w\ell H$. If the superfluid exerts a force $\vec{F}^{(M)}/H$ per unit length on each vortex, then every vortex must exert an equal and opposite force on the superfluid. Let $\vec{f}_{ns}$ be the mutual friction force per unit volume experienced by the superfluid due to vortices. This force is therefore equal to
\begin{equation}
    \vec{f}_{ns} = -N_L\frac{\vec{F}^{(M)}}{V_c} = -\alpha \rho_s \kappa L (\vec{v}_s - \vec{v}_n) + \alpha' \rho_s \kappa L \hat{z} \times (\vec{v}_s - \vec{v}_n),
\end{equation}
where $N_L$ is the number of vortices, and $L = N_L/w\ell$ is the vortex density per unit area. Our experiment is sensitive only to the dissipative term of the mutual friction force parallel to $\vec{v}_s$, and takes place in a channel which is sufficiently small to clamp normal fluid motion such that $\langle \vec{v}_n \rangle = 0$. We can therefore model the average frictional force per unit volume felt by the superfluid as
\begin{equation}
    \langle \vec{f}_f \rangle = -\alpha \rho_s \kappa L \langle \vec{v}_s \rangle.
    \label{eq:friction_per_volume}
\end{equation}

\section{Vortex Pinning}

Pinning of vortices can occur due to surface roughness. Because a vortex line can minimize its energy by reducing its length, the ends of the vortex line will tend to pin themselves to the peaks of protrusion, where motion in any direction will increase the vortex energy. This creates an elastic pinning force that opposes the Magnus force that drives motion in the vortex line. If the displacement of the vortex line becomes too great, then it may de-pin and jump to the next available pinning. For a vortex of length $H$ pinned to a hemispherical protrusion of size $b$, Schwarz \cite{schwarz1985} estimates the depinning velocity to be 
\begin{equation}
    v_d = \frac{\kappa}{2\pi H} \ln\left(\frac{b}{a_0}\right),
\end{equation}
where $a_0 \approx \xi_0$ is the vortex core size. This equation is expected to hold as long as $a_0 \ll b$, which has been demonstrated in $^4$He to a high degree of accuracy for similar nanofluidic devices \cite{varga2025}. In our system, however, the surface roughness is expected to be on the order of nanometers, whereas the vortex core size in on the order of tens of nanometers. In this case, we expect the pinning to be dominated by inter-vortex interaction instead \cite{sonin1992}. The de-pinning velocity predicted by Sonin \cite{sonin1992} for collective pinning is
\begin{equation}
    v_d = \frac{\kappa}{2\pi H} \ln\left(\frac{r_v}{a_0}\right),
\end{equation}
where $r_v$ is the average vortex separation. Here we assume the aspect ratio of the protrusion to be $1$, in keeping with the assumption of hemispherical pinning sites by Schwarz. In our case where $r_v \approx 2H$, and $a_0 \approx \xi_0$, then $\ln(2H/\xi_0) \approx 1.2-2.0$ for all devices. The de-pining velocities then are $9.7-14.6$ mm/s for the 636 nm device, $8.3-12.2$ mm/s for the 805 nm device, and $6.8-9.7$ mm/s for the 1067 nm device. In all cases, the de-pinning velocity exceeds the critical velocity at which non-linear effects become apparent in our devices. It is worth noting that these estimates of collective pinning are based on Sonin's calculations, which assume vortices with circulation in the same direction. An ensemble of vortices with random circulation directions may in fact have a higher de-pinning velocity. Vortices are therefore expected to remain pinned. This does not mean the vortices are static, however. The vortex lines may still move collectively in response to the Magnus force while maintaining the same mean separation. Support for such a theory of mutual friction with collective pinning has been demonstrated in superfluid $^3$He \cite{sonin1993,sonin1993collective}.

\section{Helmholtz Resonator Calibration}

Here, we consider the effect of vortex mutual friction on broadening the observed Helmholtz mode. The equation of motion for fluid in the channels is 
\begin{equation}
    \ddot{y} + \frac{F_f}{m_s} + \omega_0^2y = \frac{F_{\scriptsize\textrm{eff}}}{m_s},
\end{equation}
where $y$ is the fluid displacement, $F_f$ is a frictional force assumed to be linear in $\dot{y}$, $\omega_0$ is the resonant frequency, $F_{\scriptsize\textrm{eff}}$ is an effective driving force, and $m_s$ is the superfluid mass given by
\begin{equation}
    m_s = 2\rho_s a \ell.
\end{equation}
Here, $a$ is the cross-sectional area of the channel and $\ell$ is the channel length. The factor of 2 comes from the presence of two channels in our geometry \cite{Shook2020}. The frictional force experienced by the resonator is connected to the mutual friction force by
\begin{equation}
    \vec{F}_f = 2  a \ell \left(\frac{\rho_s}{\rho} \right) \langle \vec{f}_{f} \rangle.
\end{equation}
Since the damping is linear, the full-width half-max is given by
\begin{equation}
    2\pi \Delta f = \frac{| \vec{F}_f |}{m_s\dot{y}} = \frac{|\langle \vec{f}_{f} \rangle|}{\rho \dot{y}}.
\end{equation}
Substituting in equation \ref{eq:friction_per_volume} and replacing $\dot{y}$ with $\langle v_s \rangle$ then gives
\begin{equation}
    \Delta f = \frac{1}{2\pi} \left(\frac{\rho_s}{\rho} \right) \alpha \kappa L.
\end{equation}
The superfluid fraction can be calibrated from the resonant frequency of the device
\begin{equation}
    \frac{\rho_s}{\rho} = (2\pi f_0)^2 \left(\frac{A^2\ell \rho}{2a}\right)\left(\frac{1+\Sigma}{k_p}\right),
\end{equation}
where $k_p$ is the stiffness of the quartz plates, and $\Sigma$ is a ratio of this stiffness to an effective $^3$He stiffness associated with its compressibility, $\chi$, such that
\begin{equation}
    \Sigma = \frac{V\chi k_p}{A^2}.
\end{equation}
Putting this together yields an equation for $L$ in terms of two resonance parameters ($f_0$ and $\Delta f$), and the mutual friction parameter $\alpha$. The full expression is
\begin{equation}
    L = \frac{1}{2\pi \kappa \alpha} \left(\frac{2a}{A^2\ell \rho} \right) \left(\frac{k_p}{1+\Sigma} \right) \frac{\Delta f}{f_0^2}.
\end{equation}

To compute the mutual friction parameter, we make use of a theoretical model successfully employed by Bevan et.~al \cite{Bevan1997,kopnin1995spectral,Kopnin1995} to fit the mutual friction of $^3$He,
\begin{equation}
    \alpha(T) = \frac{1}{a_K} \left(\frac{\rho_s}{\rho} \right) \left(\frac{\Delta}{k_BT} \right)^{-2} e^{-\Delta/k_BT} \coth\left(\frac{\Delta}{2k_BT} \right).
    \label{eq:alpha_model}
\end{equation}
\noindent The value $a_K$ is an empirical fitting factor and $\Delta$ is the superfluid gap. This expression is explicitly derived for the bulk B-phase, however single-phase vortices in the A-phase are found to be also well fit by this expression \cite{Bevan1997}. Theoretical analysis of the mutual friction parameter under high confinement has not been carried out to date, so we assume the functional form to be the same as the bulk expression, noting that the temperature dependence closely matches our observations. Since $\rho_s$ is known to be suppressed in highly confined $^3$He, we opt to use the $\rho_s$ calibrated from each device, rather than the bulk curve. This means our calculation of $\alpha$ does, in an indirect way, include some information about the confinement. Since $\Delta$ is not directly accessible to us via experiment, we use the bulk $B$-phase BCS curve. The weak coupling gap can be calculated using the equation
\begin{equation}
    \ln\left(\frac{T}{T_c}\right) = \frac{2\pi k_B T}{\Delta_B(T)} \sum_{n=0}^{\infty} \left( \frac{1}{\sqrt{1+x_n^2}} - \frac{1}{x_n} \right),
\end{equation}
where 
\begin{equation}
    x_n(T) = (2n+1) \frac{\pi k_BT}{\Delta_B(T)},
\end{equation}
are the normalized Matsubara frequencies. Because the terms fall off for large $n$, we truncate the sum at $n=1000$. We follow the so-called weak-coupling-plus method for adjusting the gap to the ``trivial" strong coupling corrections. This is achieved by rescaling the temperature $T \to T_s$ such that 
\begin{equation}
    \frac{\Delta_B(T_s)}{k_BT_s} = \kappa \frac{\Delta_B(T)}{{k_BT}}.
    \label{eq:Ts_eq}
\end{equation}
Here $\kappa$ is an empirical parameter derived from the $A-B$ heat capacity jump
\begin{equation}
    \kappa = \sqrt{\frac{1}{1.426}\frac{\Delta C}{C}},
\end{equation}
which contains information about the strong coupling. The temperature rescaling is achieved by numerically iterating $T_s$ until it satisfies equation \ref{eq:Ts_eq}.
\begin{figure}
    \centering
    \includegraphics[width=0.5\linewidth]{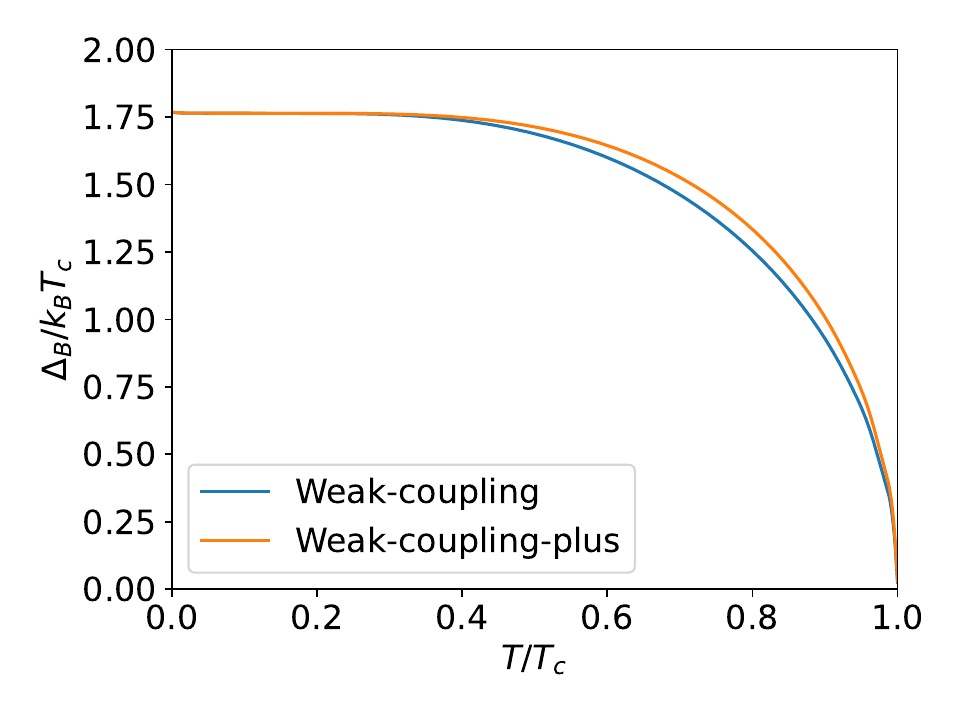}
    \caption{Plot of the BCS gap, with and without the ``trivial" strong coupling corrections.}
    \label{fig:BCS_Gap}
\end{figure}

\begin{figure}
    \centering
    \includegraphics[width=\linewidth]{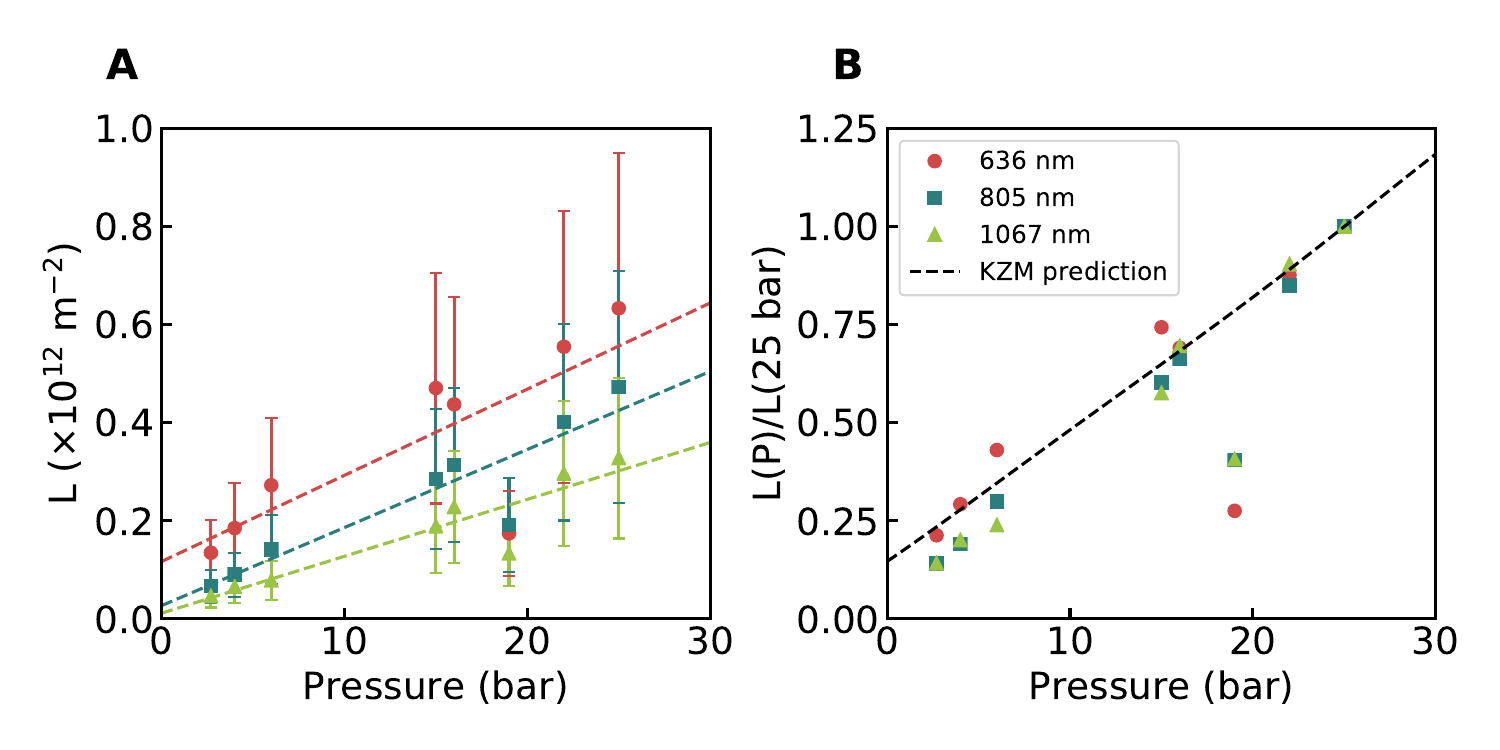}
    \caption{\textbf{Pressure Dependence of Vortex Density.} \textbf{(a)} A linearly scaled plot of the vortex density as a function of pressure. Here, the dashed lines are fits to the data. A similar linear dependence is seen in each device. \textbf{(b)} The pressure dependence of the data is compared to the Kibble-Zurek prediction, by normalizing the vortex densities to the value at 25 bar. Although the KZM fails to predict the density of vortices, the pressure scaling appears to be similar.}
    \label{fig:linear_pressure_dependence}
\end{figure}

In the calculation of $\alpha$, we note that Bevan et.~al \cite{Bevan1997} have reported the fitting factor $a_K$ to be in the range $0.02$-$0.06$ depending on both pressure and temperature. It is unclear to what extent this can be generalized beyond the experiment of Bevan et.~al, especially since a secondary fit factor is introduced into the model \cite{Bevan1997}. For the sake of keeping our model simple, we assume $a_K  \approx 0.04 \pm 0.02$, independent of temperature and pressure. This is propagated into the error bars on the vortex density shown in Fig.~\ref{fig:linear_pressure_dependence} and in the main text.

\begin{figure}
    \centering
    \includegraphics[width=\linewidth]{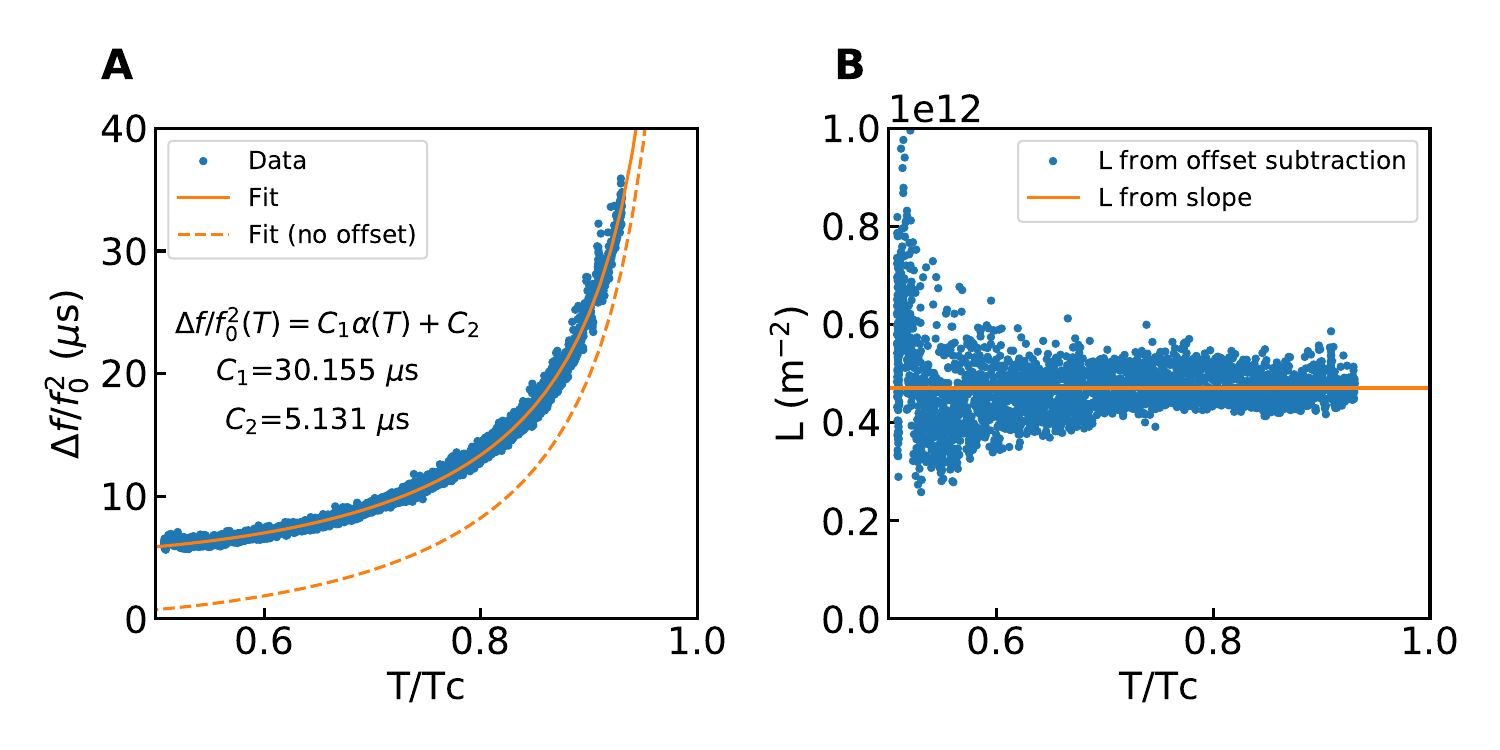}
    \caption{\textbf{Vortex Density Calibration Fit.} \textbf{(a)} Plot of the experimental ratio $\Delta f/f_0^2$ as a function of temperature at 15 bar for the 636 nm device. This data has been fit to our model and plotted as a solid orange line. The dashed line displays the same curve if the offset is omitted in order to highlight the modification. \textbf{(b)} The vortex density is computed using the two methods shown in equation \ref{eq:L_two_meathods}. Blue points are obtained by subtracting the fit offset $C_2$ from the data and dividing out pre-factors at each temperature. The orange line is computed directly from the slope $C_1$. The two methods are found to be consistent.}
    \label{fig:alpha_fit}
\end{figure}

\begin{figure}
    \centering
    \includegraphics[width=\linewidth]{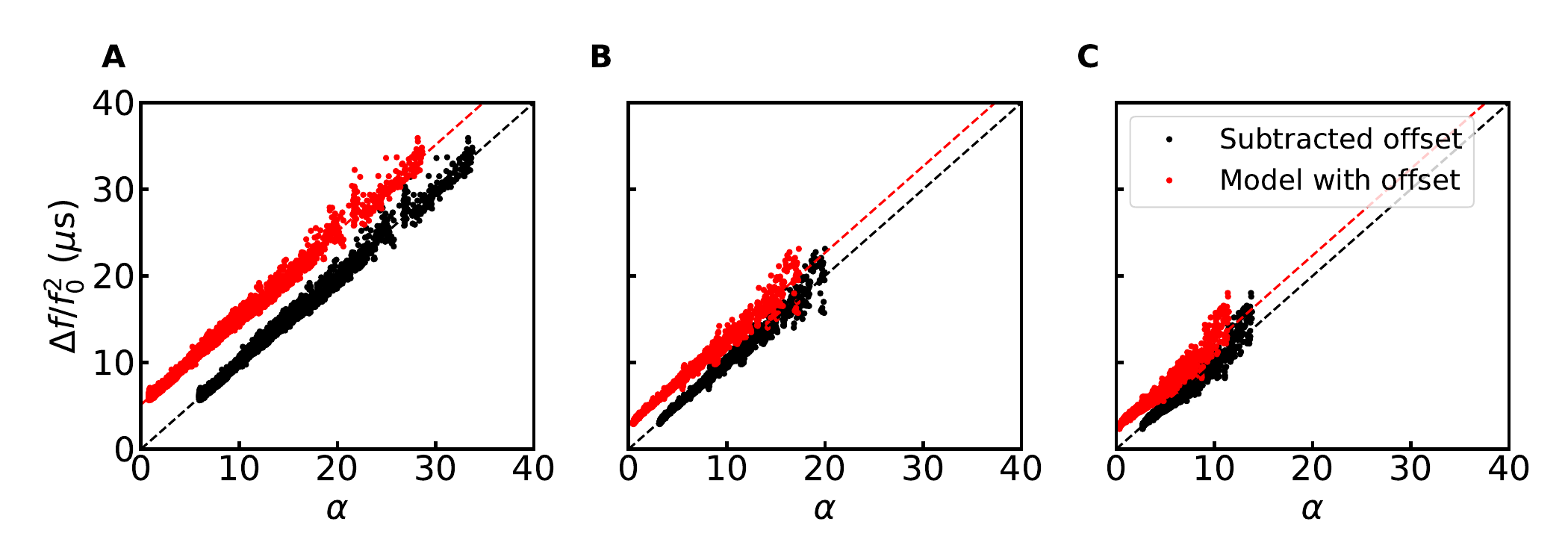}
    \caption{\textbf{Linearity in Mutual Friction Parameter.} Plots of sample data for the \textbf{(a)} 636 nm, \textbf{(b)} 805 nm, and \textbf{(c)} 1067 nm devices taken at 15 bar. The experimental ratio $\Delta f/f_0^2$ is plotted against the theoretically derived $\alpha$ in red in order to highlight the linear dependence. Extrapolation to $\alpha=0$ makes it clear that a constant offset exists in the data. Points in black show the same data with this constant offset term subtracted for the purpose of comparison.}
    \label{fig:Vortex_Density_Linear_Fit}
\end{figure}

If the vortex density $L$ is independent of temperature (as is the case for remanent vortices), then our model predicts that the measured ratio $\Delta f/f_0^2$ should have the same temperature dependence as $\alpha$. Comparison of the data to equation \ref{eq:alpha_model} shows excellent agreement, Fig.~\ref{fig:alpha_fit}, but with a small constant offset between the two curves. To account for this discrepancy we modify the model to include an empirical offset
\begin{equation}
    \frac{\Delta f(T)}{f_0^2(T)} = C_1 \alpha(T) + C_2
    \label{eq:offset_model}
\end{equation}
where
\begin{equation}
    C_1 = 2\pi \kappa L \left(\frac{A^2 \ell \rho}{2a} \right) \left(\frac{1+\Sigma}{k_p} \right) = \gamma L,
    \label{eq:C1}
\end{equation}
and $C_2$ is a fitting constant with units of time. 
Fig.~\ref{fig:Vortex_Density_Linear_Fit} plots the observed ratio $\Delta f/f_0^2$ against $\alpha$, as computed from equation \ref{eq:alpha_model}, making it obvious that the relationship is linear with a constant offset. The vortex density can then be extracted from either the slope or offset fit constants
\begin{equation}
    L = \frac{C_1}{\gamma} = \frac{(\Delta f/f_0^2-C_2)}{\gamma \alpha}.
    \label{eq:L_two_meathods}
\end{equation}
A comparison of the two methods can be seen in Fig.~\ref{fig:alpha_fit}. Elsewhere, the slope method is used wherever the vortex density is quoted. 

\begin{figure}
    \centering
    \includegraphics[width=\linewidth]{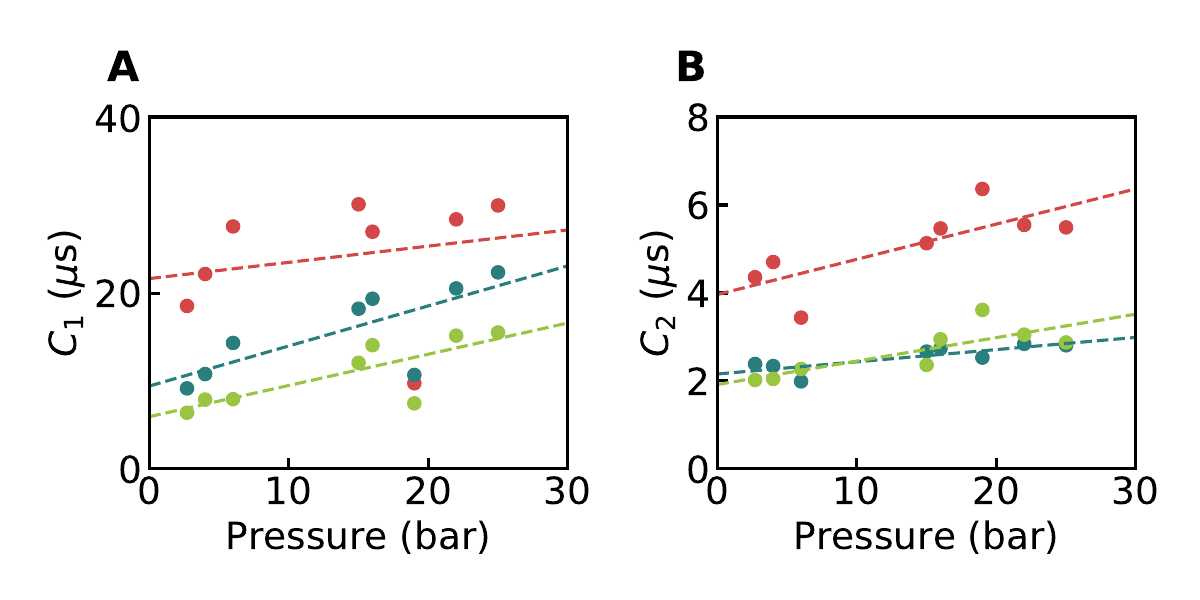}
    \caption{\textbf{Fitting Constants.} Plots showing the pressure dependence of the \textbf{(a)} slope and \textbf{(b)} constant offset for our model of $\Delta f/f_0^2$. The dashed lines represent fits to the data.}
    \label{fig:fit_constants}
\end{figure}

The fact that the temperature dependence of the experimental ratio $\Delta f/f_0^2$ is very well modeled by a function linear in $\alpha(T)$ is taken to be a good indication of the validity of our assumptions: namely that the dissipation is in fact dominated by vortex mutual friction created by dynamic vortices, where the density sets in after the transition and does not change appreciably throughout the experiment. The physical origins of the constant $C_2$ are not explored in this publication. It is possibly a simple discrepancy in the mutual friction model of the form $\alpha(T) \to \alpha(T) +$constant, or an additional effect not accounted for in our model. Regardless, the presence of a constant offset is not expected to influence the accuracy of our derived $L$, since only forces proportional to $\alpha(T)\rho_s(T)$ are identified as due to vortex mutual friction in our model.

\section{Kibble Zurek Mechanism}

A phase transition may be parameterized in terms of a dimensionless distance \cite{Zurek1985}, $\varepsilon$, from the transition temperature, $T_c$, which is assumed to change approximately linearly in time such that
\begin{equation}
    \varepsilon = 1 - \frac{T}{T_c} \approx \frac{t}{\tau_Q}.
    \label{eq:epsilon}
\end{equation}
Here the time, $t$, is normalized by a quench time, $\tau_Q$, defined by the temperature ramp rate and critical temperature
\begin{equation}
    \tau_Q = T_c\left(\frac{dT}{dt}\right)^{-1}.
\end{equation}
The equilibrium correlation length and relaxation time both diverge near the transition temperature such that
\begin{equation}
    \xi = \xi_0 \varepsilon^{-\nu},
\end{equation}
\begin{equation}
    \tau = \tau_0 \varepsilon^{-z\nu},
    \label{eq:RelaxationTime}
\end{equation}
where $\nu$ and $z$ are critical exponents specific to the system \cite{Zurek1985}. In the context of $^{3}$He, $\nu=1/2$ and $z=2$. The relaxation time, $\tau$, represents a time required for the system to reach equilibrium. Initially, the relaxation time will be sufficiently small for the system to evolve adiabatically, but due to the divergence at $\varepsilon=0$, there will reach a point where the evolution of the order parameter can no longer keep up with changes in $\varepsilon$ \cite{Campo2014}. Kibble-Zurek theory treats this problem by dividing the phase transition into a pre-transition adiabatic stage, an intermediate `impulse' stage, and a post-transition adiabatic stage \cite{Campo2014}. During the adiabatic stages, where the relaxation time changes slowly, it is assumed that there is no time delay between $\varepsilon$ and the order parameter. During the impulse stage, where the relaxation time rapidly diverges, the microscopic degrees of freedom are assumed to be frozen in place. The time at which the system enters the impulse stage, referred to as the freeze-out time, $\hat{t}$, is defined as the point before the transition at which the relaxation time is equal to the total remaining time until the transition occurs $\hat{t} = \tau(\hat{t})$. Using equations \ref{eq:epsilon} and \ref{eq:RelaxationTime}, this yields the solution
\begin{equation}
    \hat{t} = (\tau_0\tau_Q^{z\nu})^{\frac{1}{1+z\nu}}.
\end{equation}
Since the system does not evolve appreciably during the impulse stage, order parameter domains become frozen in, with an average size of
\begin{equation}
    \hat{\xi} = \xi(\hat{t}) = \xi_0 \left(\frac{\tau_Q}{\tau_0} \right)^{\frac{\nu}{1+z\nu}}.
\end{equation}
This frozen-in domain size $\hat{\xi}$ is used to approximate the density of defects to be 
\begin{equation}
    L = \frac{1}{\hat{\xi}^{2}} =  \frac{1}{\xi_0^2} \left(\frac{\tau_0}{\tau_Q} \right)^{\frac{2\nu}{1+z\nu}}.
\end{equation}

In $^3$He, the Ginzburg Landau coherence length is a function of $T_c$ and the Fermi velocity, $v_F$ \cite{Vollhardt2013},
\begin{equation}
    \xi_0 = \frac{\hbar v_F}{2\pi k_B T_c},
\end{equation}
and we take the relevant velocity scale for setting the coherence time to be the Fermi velocity \cite{Bauerle1996} such that
\begin{equation}
    \tau_0 = \frac{\xi_0}{v_F} = \frac{\hbar}{2\pi k_B T_c}.
\end{equation}
We obtain values for the Fermi velocity and critical temperature at each pressure using fit coefficients from \cite{Halperin1990}. The predicted vortex density according to Kibble-Zurek theory in a $^3$He system therefore is
\begin{equation}
    L = \frac{1}{\xi_0^2}  \sqrt{\frac{\xi_0}{v_F T_c} \frac{dT}{dt}} = \frac{1}{v_F^2} \left(\frac{2\pi k_B T_c}{\hbar}\right)^{3/2} \sqrt{\frac{1}{T_c}\frac{dT}{dt}}.
\end{equation}

It is worth noting that in systems with reduced dimensionality, the critical exponents may differ. Our system occupies an intermediate regime where the Ginzburg Landau coherence length is small compared to the separation of the walls, but the Kibble-Zurek length scale, $\hat{\xi}$ is large. We cannot directly measure how this might change in critical exponents, however, we can compare to the theoretical work of Gluscevich and Sauls \cite{gluscevich2024kibble}, who study the formation of domain walls in $^3$He using a time-dependent Ginzburg Landau simulation. They find that the value of the exponent $\beta = 2\nu/(1+z\nu)$ differs from the mean-field value of $\beta = 1/2$. When calculating the density of defects, their simulation results are best fit by the value $\beta = 0.517$. We note that if we were to take this modified exponent, it would only reduce the calculated vortex density by a factor of 1.26. Since we are interested in order of magnitude estimates, such a small change does not change our interpretation qualitatively.

\section{Drive Dependence}

We employ a chirped pulse excitation method to drive our devices where the RMS voltage is 50 mV \cite{Shook2020}. The return current is amplified using a transimpedance amplifier and then Fourier transformed to obtain the frequency response of each device. The amplitude of the current resonance can be shown to be proportional to the superfluid velocity inside the Helmholtz resonator channels \cite{Varga2020}.

\begin{figure}
    \centering
    \includegraphics[width=\linewidth]{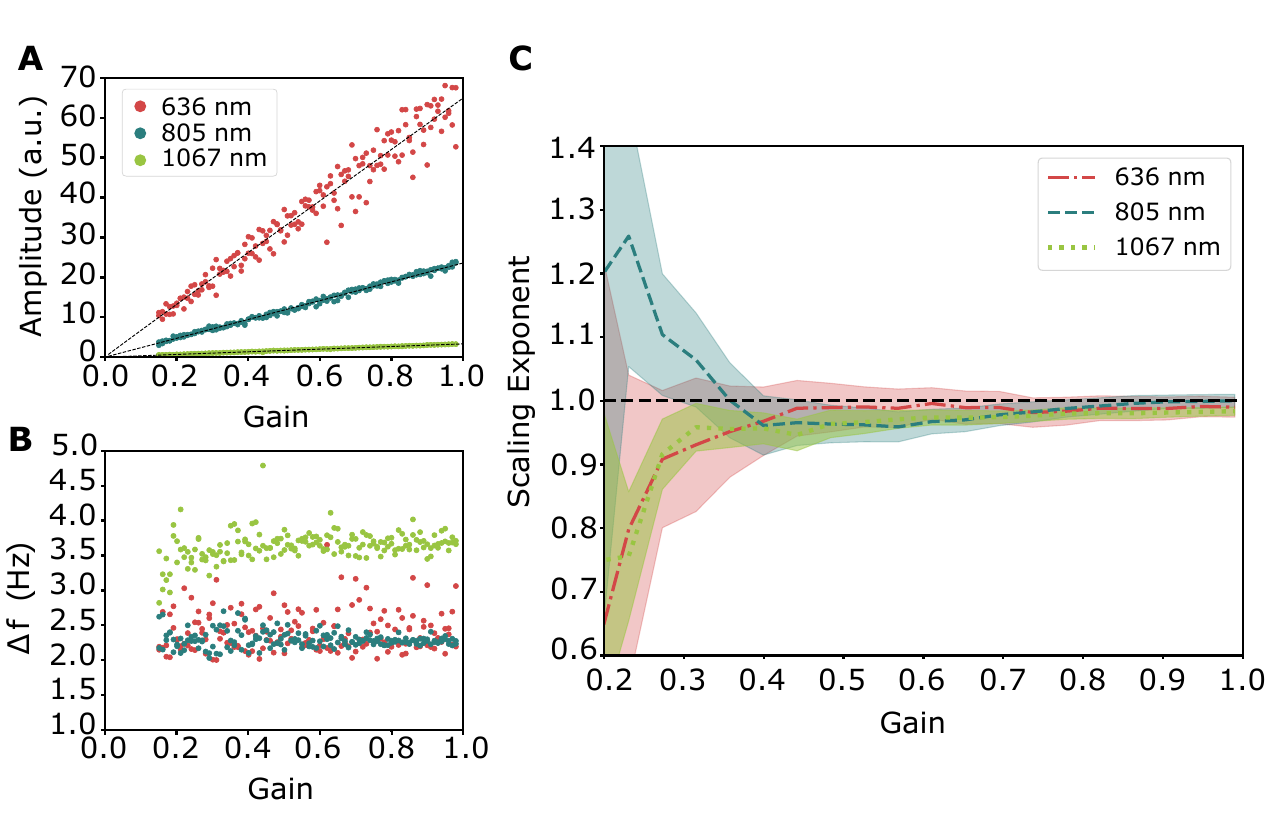}
    \caption{\textbf{Linearity Check.} \textbf{(a)} The amplitude of each Helmholtz resonator was measured while varying the drive power from a gain of 1 (corresponding to an RMS voltage of $50$ mV) to $\sim 0.2$. Past this point, the signal-to-noise ratio becomes too small for the data to be fit accurately. The data presented is from the 12.7 bar run. All other pressures are similar and exhibit a linear slope. \textbf{(b)} The linewidth of the Helmholtz resonances was observed to be independent of drive power. This is consistent with an oscillator model featuring linear dissipation. \textbf{(c)} Under the assumption that the amplitude is linear in power, the log plot should have a slope of unity. This slope, referred to as the scaling exponent, is determined through fitting the slope of the log plot to show the degree to which the data conforms to a linear dissipation model. Deviations from unity at low gain are due to the increased uncertainty (depicted by the colored regions) resulting from the decreased signal-to-noise ratio.
    }
    \label{fig:Linearity_Check}
\end{figure}

In our analysis of the vortex density within the Helmholtz resonator channels, we assume that the dissipation is linear in the average superfluid velocity $\langle v_s \rangle$, however for high drives non-linear turbulent behavior may onset \cite{Varga2020}. To verify that our experiment occupies a linear drive regime, we sweep the drive power and plot the resonance amplitude and linewidth. We note that the electrostatic force driving the motion of the quartz chips is 
\begin{equation}
    |F_{es}| = \frac{\epsilon A}{2} \left(\frac{U_D}{H}\right)^2,
\end{equation}
where $\epsilon$ is the dielectric constant of $^3$He, $A$ is the area of the electrode, and $U_D$ is the drive voltage. This electrostatic force is converted into an effective Helmholtz driving force experienced by the superfluid, which is rescaled by the following parameters
\begin{equation}
    \frac{F_{\scriptsize\textrm{eff}}}{m_s} = \frac{|F_{es}|}{\rho A \ell (1+\Sigma)}.
\end{equation}
Assuming a Lorentzian line shape, on resonance the amplitude, $y_0$, is
\begin{equation}
    y_0 = \frac{F_{\scriptsize\textrm{eff}}/m_s}{(2\pi)^2f_0\Delta f} =  \frac{1}{(2\pi)^2f_0\Delta f} \frac{|F_{es}|}{\rho A \ell (1+\Sigma)}.
\end{equation}
This implies that the product $y_0 \Delta f$ is quadratic in drive voltage, or linear in power. This is consistent with the Helmholtz resonator data, which exhibits linear power dependence in the amplitude and no power dependence in the linewidth, as shown in Fig.~\ref{fig:Linearity_Check}. The linear scaling of the amplitude is verified using a fit of the log-scaled data, which is expected to a have a slope of unity. This is found to be true except at low power where the uncertainty becomes large.

\section{Alternative Dissipation Mechanisms}

Here, we consider the possibility of Helmholtz resonator dissipation due to both non-superfluid loss mechanisms and other known mechanisms in superfluid $^3$He. Dissipation in Helmholtz resonators of a similar design has been studied in detail in the context of superfluid $^4$He \cite{Souris2017}. Loss mechanisms in this system included imperfect normal fluid clamping, heat flow from the device basin to the helium reservoir, and dissipation internal to the substrate. Loss internal to the quartz plates will freeze out well above $^3$He temperatures, meaning this can safely be ruled out. Heating-dependent dissipation is expected to be dependent on the drive power, yet we observe no appreciable change in the linewidth of our signal when sweeping the power, Fig.~\ref{fig:Linearity_Check}. This suggests that any heating generated, either by the motion of the plate or by Ohmic heating due to the electrodes, has a negligible effect.

The importance of the normal fluid in damping oscillations depends on the degree to which it is clamped, characterized by the viscous penetration depth
\begin{equation}
    \delta = \sqrt{\frac{2\eta}{\rho_n \omega}},
\end{equation}
where $\eta$ is the effective viscosity, $\rho_n$ is the normal fluid density, and $\omega$ is the angular frequency of oscillation. When the viscous penetration depth is large compared to the confinement, $\delta >> H$, the normal fluid velocity does not change appreciably across the slab. A third relevant length scale in $^3$He is the mean free path, $\ell_{\scriptsize\textrm{mfp}}$, which grows rapidly at low temperatures, and is always comparable to, or larger than, the nanoscale confinement of our experiment \cite{Einzel1997}.

When the mean free path is large compared to the confinement, the fluid is said to be in the ``Knudsen" regime and the normal fluid velocity goes to zero at the wall. If both conditions $\delta >> H$, and $\ell_{\scriptsize\textrm{mfp}} >> H$ are met, then the normal fluid is essentially completely locked. For a mean free path comparable to, or smaller than, the slab thickness, there may be a small non-zero normal fluid velocity at the wall, allowing for ``slip" of the normal fluid.

The quality factor due to normal fluid slip in this regime has been calculated using the theory of fourth sound attenuation carried out for a parallel plate geometry by Jensen et.~al \cite{Jensen1983}

\begin{equation}
    Q_s^{-1} = \frac{1}{6}\frac{\rho_n}{\rho_s} \left(\frac{H}{\delta}\right)^{2} \left[1 + \frac{12\eta}{\rho v_F H} \left(\frac{m}{m^*} \right)(e^{\Delta/k_BT}+1) \frac{(1+s)}{(1-s)} \right], 
\end{equation}
where $m^{*}/m$ is the effective mass ratio, and $1-s$ characterizes the degree of scattering specularity ($s=0$ corresponds to diffusive scattering, and $s=1$ is purely specular scattering). Theoretical calculations of the A to B transition reported in a prior publication lead us to believe the scattering in our devices is near the diffusive limit $s \sim 0$ \cite{Shook2020}. For the $636$ nm and $805$ nm devices we find $Q_s$ to be well in excess of $1000$ at all pressures, having a negligible effect on the linewidth. For the $1067$ nm device the effect becomes more applicable, accounting for up to $\sim 10 \%$ of the observed linewidth at the highest pressures.

The role of dissipation via thermoviscous effects is also considered. Motion of the superfluid, with a clamped normal fluid generates a small temperature gradient between the basin and the external helium reservoir \cite{Souris2017}. The resulting heat flow then generates loss, as well as a fountain effect which opposes the motion of the superfluid. We model the quality factor due to this effect as
\begin{equation}
    Q_{\scriptsize\textrm{th}} = \frac{1+\Phi_{\scriptsize\textrm{th}}^2}{\Phi_{\scriptsize\textrm{th}}\Gamma^2} \left(1+\frac{\Gamma_{\scriptsize\textrm{th}}^2}{2(1+\Phi_{\scriptsize\textrm{th}}^2)} \right).
\end{equation}
The first dimensionless quantity $\Phi_{\scriptsize\textrm{th}}$, is defined as
\begin{equation}
    \Phi_{\scriptsize\textrm{th}} = \frac{1}{\omega_0 \tau_{\scriptsize\textrm{th}}},
\end{equation}
which compares the angular frequency of the Helmholtz resonance to the thermal relaxation time, $\tau_{\scriptsize\textrm{th}}$, required for the basin to equilibrate with the reservoir. This time constant is give by the product of the heat capacity of the basin $C_{\scriptsize\textrm{th}}$ and an effective thermal resistance, $R_{\scriptsize\textrm{th}}$, between the two volumes, $\tau_{\scriptsize\textrm{th}} = C_{\scriptsize\textrm{th}}R_{\scriptsize\textrm{th}}$. The second quantity, $\Gamma_{\scriptsize\textrm{th}}$, compares the reactive force generated by the fountain effect to the effective stiffness of the Helmholtz resonator
\begin{equation}
    \Gamma_{\scriptsize\textrm{th}}^2 = \left(\frac{A_BS}{V_B}\right)^2 \frac{T}{C_{\scriptsize\textrm{th}}} \frac{1+\Sigma}{k_p/2}.
\end{equation}
Here, $A_B$ is the area of the basin, $V_B$ its volume, and $S$ the total entropy of the fluid in the basin. Based on a calculation found in the supplementary material of Ref.~\cite{Souris2017}, we estimate $R_{\scriptsize\textrm{th}} \sim 9.85 \times 10^3$ K/W. The superfluid heat capacity and entropy are computed using numerical equations from \cite{Carless1983}. For the range of temperatures, pressures and confinements relevant to this experiment, we therefore estimate $Q_{\scriptsize\textrm{th}} \sim 10^{3}-10^{4}$, corresponding to line widths on the order of $\sim 0.01-0.1$ Hz.

Lastly, non-remanent vortices may form through mechanisms such as rotational flow, or turbulence. Simulations of the Helmholtz resonator devices predict a small amount of vorticity to be created by superflow around the corners of the channel, as described in Ref.~\cite{Varga2020,shook2024surface}. Because this effect is expected to be drive-dependent \cite{Varga2020} contrary to our observations, this is not compatible with the measurement.

\section{Vortex Recombination}

\begin{figure}
    \centering
    \includegraphics[width=\linewidth]{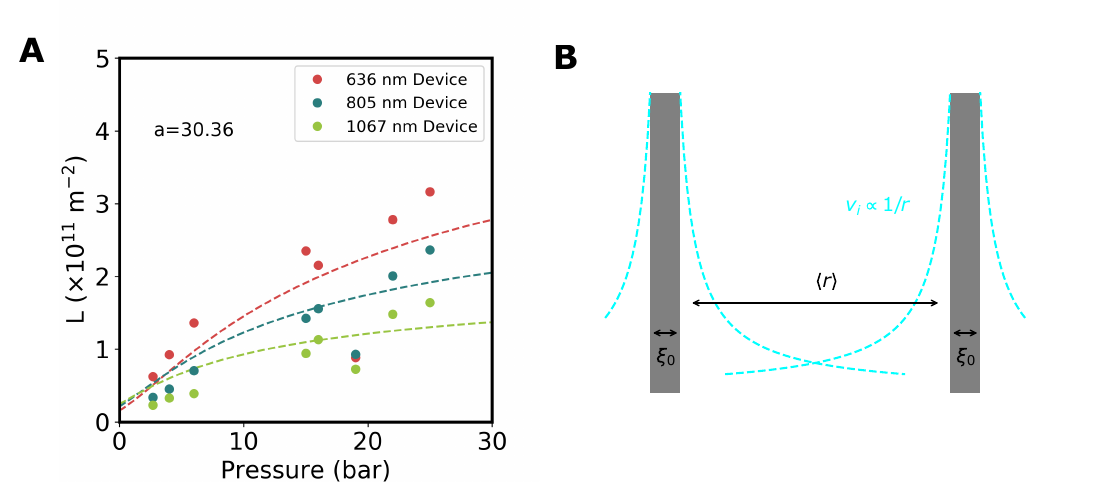}
    \caption{\textbf{(a)} Pressure dependence of vortex density for each channel confinement plotted on a linear scale. We find that all three data sets can be fit by the function $L(H,P) = \exp(-a\xi_0(P)/H)/(2H)^2$, where $a=30.36$ is a pressure independent parameter common to all three devices. We propose that this may be due to the fact that the vortex core size scales with $\xi(P)$, which increases the probability of vortex recombination at low pressures. \textbf{(b)} Diagram of vortex lines (grey), with core sizes of $\xi_0$, separated by an average distance of $\langle r \rangle \sim 2H$. The velocity field induced by each vortex $v_i$ is shown in cyan to fall off as $1/r$ with some overlap between the two vortices. Vortices that form closed enough together may recombined due to interactions. Our model suggests that vortices initially form with an average separation $\langle r \rangle \approx 2H$, which is 10-100 core diameters depending on pressure. The defect density may decrease (increasing average separation) due to vortex recombination shortly after the phase transition.}
    \label{fig:vortex_depletion}
\end{figure}

We have seen that the density of vortices in each device is predicted to within an order of magnitude by $L \sim (2H)^{-2}$. This however does not account for the pressure dependence observed in our measurements. A pressure dependent length scale that might modify the density of vortices is the vortex core size, which is on the order of the coherence length $\xi_0 = 20-80$ nm. The average spacing between vortices, meanwhile is
\begin{equation}
    \langle r \rangle = \frac{1}{\sqrt{L}} \approx 2H.
\end{equation}
This means the average vortex separation is expected to be 10-100 core diameters depending on the pressure. Since the velocity induced by a vortex falls off as $1/r$, interaction between vortices at these separations is not insignificant, and vortex annihilation may occur shortly after the phase transition. This would mean that more vortices would annihilate at the lowest pressures where the core size is largest. We find that the pressure dependence can be fit as a function of $\xi_0/H$ with negative exponential dependence
\begin{equation}
    L = L_{\mathrm{tot}}e^{- a \left( \xi_0/H \right)}, \qquad L_{\mathrm{tot}} = \frac{1}{(2H)^2},
\end{equation}
where $a$ is a fitting constant that is independent of channel confinement and pressure. We fit this model to the pressure dependence of $L$ for each device (see Figure \ref{fig:vortex_depletion}), and find that all three data sets are well fit by a single free parameter $a = 30.3$.

\section{Ramp Rate Dependence}

Ramp rate measurements were performed in two different ways; continuous temperature ramps, and discrete temperature steps. In the continuous method, the Helmholtz resonator was continuously probed using excitation pulses. The time-domain data from the capacitance bridge is Fourier transformed to extract information about the resonance. To achieve a satisfactory signal-to-noise ratio we typically average 10-20 pulses per data point, with a pulse spacing of $\sim$ 1 second or less. The typical ramp rate in most experiments was 0.12 mK/hr. For the second method, we ramp the temperature at a given rate to the desired temperature and wait for the system to equilibrate before taking each resonance measurement.

\begin{figure}
    \centering
    \includegraphics[width=0.5\linewidth]{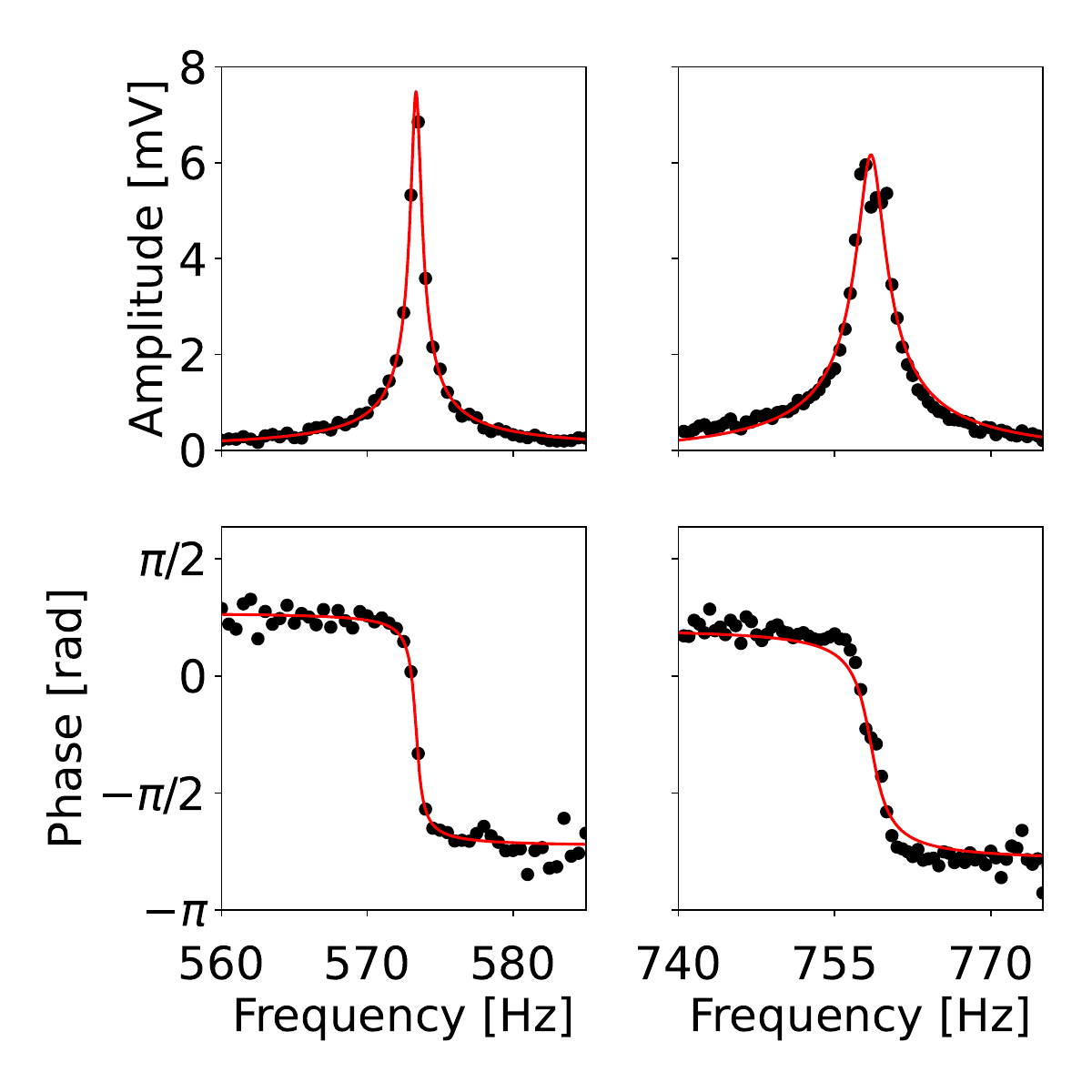}
    \caption{Examples of the amplitude and phase for a good Lorentzian fit, measured at a stationary temperature, and a bad fit broadened due to averaging.}
    \label{fig:Distorted_peak}
\end{figure}

\begin{figure}
    \centering
    \includegraphics[width=\linewidth]{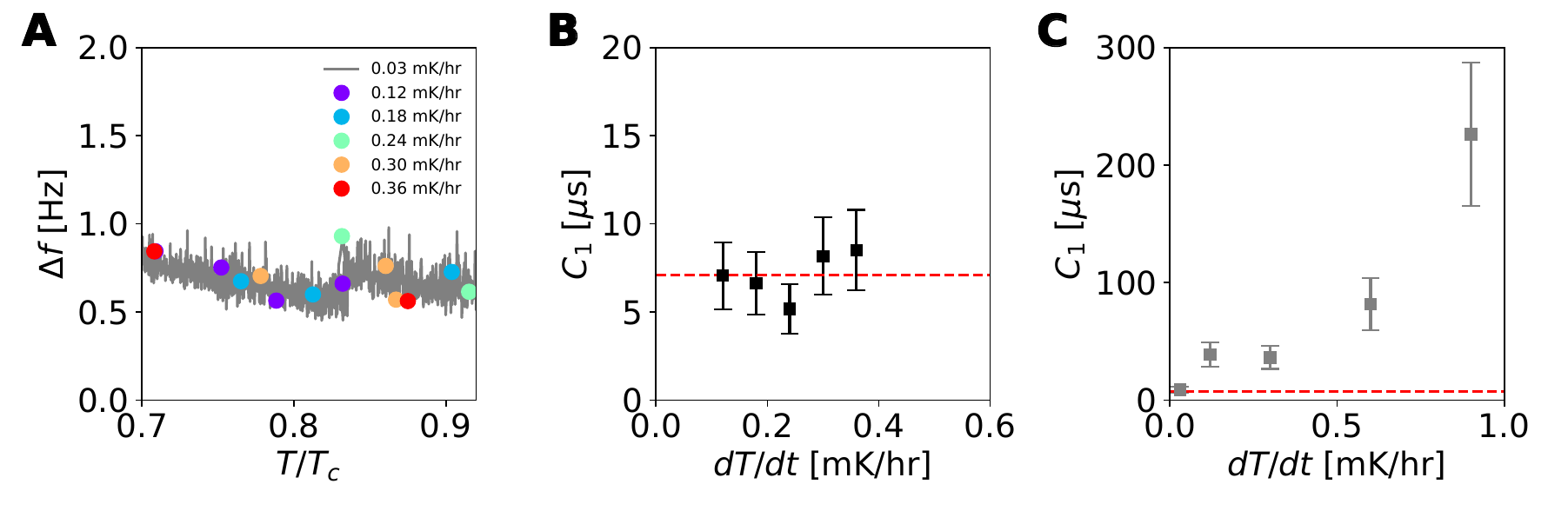}
    \caption{(a) Plot of the line width measured in the 750 nm device at 14.3 bar. The grey line shows the linewidth measured when the temperature is ramped continuously at 0.03 mK/min. The colored points show the linewidth measured at faster ramp rates (0.12-0.36 mK/min) where the measurement temperature ramp is stopped at discrete temperature values. When the ramp rate is slow for the continuous measurement, there is agreement between the two methods. (b) Plot of the fitting constant $C_1$ (proportional to vortex density $L$) as a function of temperature ramp rate for the discrete temperature step measurements. The dashed red line shows the mean value. (c) Plot of $C_1$ for the continuous measurement method. Here the dashed red line again shows the mean value of the discrete temperature step method for comparison. At 0.03 mK/min this agrees with the temperature step method. At intermediate ramp rates of $>$ 0.03 mK/hr and $<$ 0.4 mK/hr there an increase in $C_1$ by a factor of 3.9 associated with the artificial broadening. For fast ramp rates $>$ 0.4 mK/hr we see a dramatic increase in $C_1$ due to the system no longer being in thermal equilibrium.}
    \label{fig:Ramp_Rate}
\end{figure}

It was discovered that this continuous measurement method at 0.12 mk/hr introduces a systematic error into the experiment. If the resonant frequency drifts appreciably over the averaging period, it can result in an artificial broadening of the peak. This becomes evident in the lineshape of the peak, which deviates from a true Lorentzian; as shown in Figure \ref{fig:Distorted_peak}.  Additionally, for sufficiently large ramp rates ($> 0.4$ mK/hr) we see evidence of thermal gradients between the experimental cell, and the lowest temperature stage of our cryostat. If the ramp rate is reduced to 0.03 A/hr or slower we find that the continuous measurement method recovers the results of the temperature step method. This is shown in Figure \ref{fig:Ramp_Rate} A. 

Since the relevant parameter in the KZM is the temperature ramp rate during the phase transition, the method of temperature steps should not result in any change to the vortex density. In Figure \ref{fig:Ramp_Rate} B we show that changing the ramp rate from 0.12-0.36 mk/hr results in no change to the fit parameter $C_1$. Figure \ref{fig:Ramp_Rate} shows $C_1$ for the continuous measurement method, which agrees with the step method only at the slowest ramp rate of 0.03 mK/hr. The difference between the two measurement methods at 1.2 mK/hr of $\sim$4 was used to quantify the degree of over-estimation in the vortex density due to artificial broadening.

From these results, we conclude that there is no change in vortex density for ramp rates from 0.03 to 0.36 mK/hr, contrary to the prediction of the KZM of $\sqrt{dT/dt}$ scaling. By varying the temperature ramp rate, we can also address a second question relevant to the nucleation of vortices. It is thought that during the phase transition from the normal fluid to superfluid phase the existence of temperature or pressure gradients may modify the vortex nucleation sites that lead to remanent vortices \cite{del2014universality}. This effect was noted in the supplementary material of \cite{rysti2021suppressing}. Temperature gradients could conceivably exist across our sample during the phase transition due to the finite thermal conductance of $^3$He. If this effect significantly modified the resulting density of defects, we would expect to see a temperature ramp rate dependence. 

To address the question of pressure gradients existing during the phase transition, we performed control experiments where the fluid driving force was applied during the phase transition, and experiments where it was not applied until after the phase transition. We see no difference between these measurements. This is in line with the results of \cite{rysti2021suppressing}, where the system was rotated in a rotating cryostat during the phase transition. The authors demonstrate that the density of vortices attributed to the KZM is independent of the rotational speed during phase transition. This appears to be consistent with our observation that the applied flow during the phase transition does not change the density of defects.

\newpage
